\documentclass[A4paper,useAMS,usenatbib]{mn2e}
\usepackage[pdftex]{graphicx}
\usepackage{color}
\usepackage[normalem]{ulem}

\def\be{\begin{equation}}
\def\ee{\end{equation}}
\def\bea{\begin{eqnarray}}
\def\eea{\end{eqnarray}}

\newcommand{\htwo}{${\rm H_2}$}

\newcommand{\ha}{H$\alpha$}
\newcommand{\chianti}{{\it CHIANTI\/}}
\newcommand{\planck}{{\it Planck\/}}
\newcommand{\herschel}{{\it Herschel\/}}

\title[Character of the WIM]{The character of the warm ionised medium}
\author[Geyer \&\ Walker]{Marisa Geyer,$^{\!\!1,2}$\thanks{E-mail: Marisa.Geyer@gmail.com}
Mark$\!$ A.$\!$ Walker$^{3}$\thanks{Mark.Walker@manlyastrophysics.org}\\ 
$\!$1. Oxford Astrophysics, Keble Road, Oxford, OX1 3RH, England\\
$\!$2. SKA South Africa, Park Road, Pinelands 7405, South Africa\\
$\!$3. Manly Astrophysics, 15/41-42 East Esplanade,  Manly, NSW 2095, Australia}

\begin{document}

\date{Accepted: 17th August 2018.\ \ \ \  Received: 17th August 2018.\ \ \ \  In original form: 18th May 2018}

\pagerange{\pageref{firstpage}--\pageref{lastpage}} \pubyear{2018}

\maketitle

\label{firstpage}

\begin{abstract}
\herschel\ observations of far infrared N$^+$ emission lines have demonstrated that dense plasma, with $n_e\sim30\,{\rm cm^{-3}}$, is ubiquitous in the inner Galactic plane. By combining the information from \herschel\ with other tracers of ionised gas, we build a picture of this dense plasma. We adopt a collisional ionisation model, so the analysis is not tied to a specific energisation mechanism. We find that the dense plasma is concentrated in a disk that is $130\;{\rm pc}$ thick, and makes a significant contribution to radio pulsar dispersion measures in the inner Galactic plane. The strength of the far infrared N$^+$ emission requires high temperatures in the plasma, with $T \simeq 19{,}000\,{\rm K}$ indicated both by the ratio of N$^+$ to C$^+$, and by the ratio of N$^+$ to microwave bremsstrahlung in the inner Galactic plane. This parallels the situation at high Galactic latitudes, where strong optical emission is observed from N$^+$ (and S$^+$), relative to both \ha\ and microwave bremsstrahlung, and suggests a common origin. If so, the same gas provides a natural explanation for the extreme radio-wave scattering phenomena that are sometimes observed in pulsars and quasars. We therefore propose a new picture of the warm ionised medium as seen in emission, in which the plasma is dense, hot, and localised in numerous structures of size $\sim 10^2\,{\rm AU}$ that are clustered around stars.
\end{abstract}


\section{Introduction}
Amongst the many interesting results flowing from the far-infrared (FIR) {\it Herschel\/} observatory, one of the most remarkable is the discovery that N$^+$ line emission arises predominantly from dense, ionised gas ($n_e=30\pm 20\,{\rm cm^{-3}}$), throughout the inner Galactic plane \citep[][G15, henceforth]{goldsmith2015}.  The physical context for this result is not completely clear, but the emission appears to be diffuse and widespread so the interstellar medium (ISM) seems to be implicated. Although the {\it Herschel\/} data are largely consistent with earlier findings from the COBE satellite \citep{bennett1994}, the dense plasma that is required does not yet have a place in our contemporary picture of the warm ionised medium (WIM) of the Galaxy \citep[e.g.][]{haffner2009}.

Optical line emission from the ionised ISM has been studied since the early 1970's \citep{reynolds1973}, and is dominated by the WIM. By comparing emission measures (${\rm EM}=\int n_e^2\, {\rm d}s$) from optical line studies, with dispersion measures (${\rm DM}=\int n_e\, {\rm d}s$) from radio pulsars, along particular lines of sight, \citet{reynolds1991} concluded that the WIM has a low density ($n_e\sim0.1\,{\rm cm^{-3}}$) and a high filling fraction ($f\ga0.2$). Subsequent studies have reinforced those findings \citep[e.g.][]{berkhuisenmuller2008,gaensler2008}. However, these conclusions rest on the assumption that both integrals are dominated by the same regions of space. That might not be true if the ISM manifests a broad range in $n_e$: DM might arise mainly from low-density plasma that has a high filling fraction, while simultaneously most of the observed EM could be due to dense plasma that has a low filling fraction. An additional reason for caution is that optical studies of the WIM are strongly affected by dust -- via absorption and scattering -- which complicates the interpretation of the spectra, and censors our view of the distant WIM, particularly at low Galactic latitudes. 

The WIM that is remote from the Sun might be different to that seen locally. Models of the Galactic distribution of free electrons typically use several, structurally distinct components in order to approximate the observed distribution of pulsar DMs \citep{taylorcordes1993,cordeslazio2002,yao2017} (henceforth TC93, NE2001 and YMW16, respectively). In these models the DM at high Galactic latitudes is ascribed primarily to a thick disk component of free electrons that has a scale-height of order $1\,{\rm kpc}$. At low latitudes a thin disk component (scale-height of order $0.1\,{\rm kpc}$) dominates the observed DMs towards the inner Galaxy. The models also incorporate ionised gas that is associated with the Galactic spiral arms, and some additional, smaller-scale structures (e.g. the Gum Nebula). It is possible, in principle, that one or more of these components is something other than warm plasma, but usually they are all assumed to be WIM. The possibility that these various components of the WIM might differ greatly in character is underlined by the fact that the thin disk is known to be much more effective in scattering radio waves than the thick disk (TC93).

Notionally, each of the free electron models (TC93, NE2001, YMW16) provides a unique prescription for the plasma density at every point in the Galaxy. However, as those models are designed to match pulsar DMs, primarily, they provide strong constraints on the average electron density, but the point-to-point density fluctuations are not directly constrained. In other words: where the free-electron models have a notional electron density $\bar{n}_e$, it is possible instead to have plasma of density $n_e\gg \bar{n}_e$ and filling fraction $f=\bar{n}_e/n_e\ll1$, so that the same DM results. Constraints on high density fluctuations are possible if measured EMs are simultaneously considered, but to date that approach has relied on optical emission lines and therefore relates primarily to the local WIM.

To study the emission from the thin disk and spiral arm components of the WIM, it is best to observe at wavelengths that are unaffected by dust --- the FIR and radio bands. Radio recombination line (RRL) studies of the ISM have been pursued for many years \citep[e.g.][]{anantha1985,alves2015} and have often targeted the bright, inner disk of the Galaxy. In principle, RRL studies yield powerful diagnostic information on the plasma conditions (density, temperature, velocity field), and they provided some early indications of widespread, dense ($n_e\sim{\rm few\;cm^{-3}}$) plasma in the inner disk \citep[e.g.][]{anantha1985}. However, at the plasma densities of interest to us RRLs are maser lines \citep{draine2011}, which makes them difficult to interpret. 

More recently, multi-frequency surveys of the radio sky by the {\it WMAP\/} and \planck\ satellites have yielded high signal-to-noise ratio, well sampled, all-sky maps of Galactic radio emissions, including the bremsstrahlung from ionised gas \citep{bennett2003,adam2016}. These bremsstrahlung maps immediately place strong constraints on the large-scale distribution of the WIM. They also yield information about the microphysics. It was shown by \citet{davies2006} and \citet{dobler2008} that the ratio of WMAP bremsstrahlung \citep{bennett2003} to \ha\ \citep{finkbeiner2003} is lower than expected for photoionised plasma at $8{,}000\,{\rm K}$. This situation led to suggestions that the plasma temperature might be very low \citep[$3{,}000\,{\rm K}$,][]{dobler2009}, or that time-dependence might be playing a role \citep{dong2011}. 

The FIR studies of G15 have provided fresh insight into the physics of the ionised ISM. They sampled the full range of Galactic longitudes, with 149 pointings distributed over the Galactic plane, targeting the two N$^+$ lines at $122\,{\rm\mu m}$ and $205\,{\rm\mu m}$. The resulting picture is dominated by the inner Galaxy, with positive detections coming almost exclusively from longitudes $|l|\la60^\circ$. The intensity ratio of the two N$^+$ transitions is sensitive to the electron density in the emitting regions over the range $3\la n_e({\rm cm^{-3}})\la3{,}000$. Although the densities derived by G15 do vary from position to position, the variations are modest and a key feature of the data is that the density is high across all the various sight lines where N$^+$ emission is detected. High density plasma is therefore the dominant contributor to N$^+$ FIR emission for the inner Galactic plane. Henceforth we refer to this dense gas as dense-WIM (D-WIM) without prejudice to the specific physical environment in which it arises (e.g. near massive stars, perhaps), or what fraction of the WIM is in this form.

The FIR perspective on the WIM is fundamentally challenging because the diffuse ISM is characterised, in part, by a typical pressure of $n\times T\sim3{,}000\;{\rm K\, cm^{-3}}$ \citep{jenkinstripp2011}, so that $n_e\sim0.15\,{\rm cm^{-3}}$ is expected in a diffuse, fully ionised region. If the WIM density is actually $200\times$ larger, one immediately faces difficult questions about the dynamics --- e.g. what confines the gas? Or, if it is not confined, why does it not simply expand until it reaches pressure equilibrium with the ambient medium?

Similar problems have been encountered for decades in studies of radio-wave propagation in the ISM, where electron densities $n_e\sim10^{2\pm1}\,{\rm cm^{-3}}$ have been inferred from large radio-wave scattering angles and their associated flux modulations \citep{rickett1990,rickett2011}. Various extreme radio-wave scattering (ERS) phenomena have been reported, depending on the nature of the radio source. For pulsars: multiple imaging \citep{cordeswolszczan1986,rickett1997}, and parabolic arcs in the ``secondary spectrum'' \citep{stinebring2001,cordes2006}. For quasars: extreme scattering events \citep{fiedler1987,fiedler1994,bannister2016}, and intra-day variability \citep{kedziora1997,dennettthorpedebruyn2000,bignall2003}. In some cases the effects are known to be transient, implying that the regions of high plasma density are quite limited in their spatial extent \citep{kedziora2006,lovell2008,debruynmacquart2015}. Estimated sizes range from $\sim1\;{\rm AU}$ to $\sim100\;{\rm AU}$. The incidence of ERS in compact radio quasars is low, e.g. $\sim10^{-3}$ for the most extreme cases of intra-day variability \citep{lovell2003}, but the small size of the individual scattering regions means that they must nevertheless be very common --- much more numerous than stars, for example. Thus, although the ERS phenomena are rare, the plasma structures that cause them are not.

This paper considers the properties of the Galactic D-WIM, with particular reference to the constraints that follow by combining the FIR spectroscopic information with other astrophysical data. Our aim is to answer basic questions such as: can the observed emission lines -- both optical and FIR -- all arise from D-WIM? If so, what is the likely temperature of the plasma? Can the D-WIM also explain the ERS phenomena? and the dispersion of radio pulses?

Previous studies of the properties of the WIM have often assumed that the plasma is photoionised \citep[e.g.][]{mathis1986}. We did not make that assumption, partly because there are other possibilities -- low-energy cosmic-rays, for example, are an attractive option \citep{walker2016} -- and partly because it is already known that the optical spectra of the WIM observed at high latitude \citep{haffner1999,reynolds1999} are difficult to explain using photoionised plasmas. The difficulty is that the WIM appears to be hotter than can be explained by UV photoionisation alone, so that additional heating processes must be invoked. Lacking a clear picture of the physics giving rise to the WIM, we chose a route which does not require us to specify those processes. Instead we simply characterise the plasma by its density and temperature and require that it be in collisional ionisation equilibrium. This is a logical progression from previous studies: if UV photoionisation models of the WIM require additional heating processes, why not simplify the picture by dropping the photoionisation and leave everything to the heating process? The resulting model is a new approach to describing the WIM which, even if it is not entirely satisfactory as it stands, provides a useful point of reference. To compute the ionisation equilibria for our model, and the resulting emission line characteristics, we employ the {\it CHIANTI\/} atomic database and the associated Python scripts provided in the {\it ChiantiPy\/} software package \citep{dere1997,delzanna2015}.

This paper is organised as follows. In the next section we describe the model of the D-WIM used in this paper. In section 3 we consider constraints that can be obtained directly from the \herschel\ data themselves. Sections 4 and 5 combine the information from \herschel\ with \planck\ data and with radio pulsar dispersion measures, respectively. We then consider the local D-WIM in \S6, using observations of microwave bremsstrahlung and optical line emission, and highlighting the connection to extreme radio-wave scattering phenomena. Discussion and conclusions follow in \S\S7,8.

\section{Model of the D-WIM}
\subsection{Spatial distribution}To characterise the distribution of dense plasma inferred from the FIR N$^+$ lines we adopt a simple description in which the electron density is approximately constant, with value $n_e=n_g=30\,{\rm cm^{-3}}$, within spherical clouds of radius $R$, and zero outside them. For this model the dispersion measure is just ${\rm DM}=n_g\int f {\rm d}s$, where $f$ is the fraction of space that is filled with such clouds. Similarly, the emission measure is ${\rm EM}=n_g^2\int f {\rm d}s$, so we have the useful relation ${\rm EM}=n_g{\rm DM}$. In \S4 and \S5 we relate the G15 results to radio data on EMs and DMs, respectively. 

We also make use of the geometric optical depth of the clouds, $\tau$, in connection with the incidence of extreme radio-wave scattering phenomena. The meaning of $\tau$ can best be conveyed by giving the probability that the line of sight intersects a cloud: it is $1-\exp(-\tau)$, and for $\tau\ll1$ that probability is approximately $\tau$. For the model we have specified, the optical depth can be written in terms of the dispersion measure as $\tau=3\,{\rm DM}/(4 R\, n_g)$. In conjunction with a microphysical description of the plasma, these three integrals specify model values for the measurables of interest.

\subsection{Line emission}
Whereas the bremsstrahlung continuum of a plasma can be accurately described with analytic formulae \citep[e.g.][]{rybicki1979}, keeping track of ionisation states and level populations, and the various processes that affect them, requires a numerical approach. For that task we use {\it ChiantiPy\/}  --- the {\it Python\/} implementation of the {\it CHIANTI\/} atomic database \citep{dere1997,delzanna2015}.

\chianti\  provides a pre-computed determination of the equilibrium populations in each of the various possible ionisation states, for each element, at a specified electron density and temperature. The equilibria are determined entirely from consideration of the rates at which various processes proceed amongst the thermal particle populations in the plasma. The solutions do not correspond to thermodynamic equilibrium, in which each process is balanced by its own inverse; instead, the total rate of ionisation, for a particular species, is balanced by the total rate of recombinations. Ionisation is mainly from electron impact, whereas recombination is mainly radiative (two-body) at low density and three-body at high densities. An important caveat is that \chianti\ does not currently include charge exchange reactions,\footnote{Other codes, notably CLOUDY \citep{ferland2017}, do include charge exchange processes.} and for some species they are important under the conditions of interest to us --- see Appendix A. Pre-computed equilibria are provided in \chianti\ for temperatures above $10^4\;{\rm K}$; for lower temperatures we utilised a {\it ChiantiPy\/} procedure kindly provided by Ken Dere (personal communication, 2016). 

The resulting, collisional ionisation equilibria are appropriate, for example, to the solar corona, and this picture is sometimes called the coronal approximation. Indeed, the {\it CHIANTI\/} software was originally developed for application to solar physics.  Whether or not this approximation yields an accurate description of the D-WIM remains to be seen, but it is a simple model that provides a useful point of reference. In particular it corresponds to the limiting case where the plasma is energised primarily by heating, with little or no direct ionisation contributed by the energising agent. The suitability of the model can therefore be assessed from the ratio of the heating rate to the ionisation rate, for any hypothetical energising process. In \S7 we return to this point and quantify the requirement. 

Throughout this paper we adopt the cosmic elemental abundance pattern specified in \chianti, which is as per \citet{allen1973}.

\begin{figure}
\includegraphics[width=85mm]{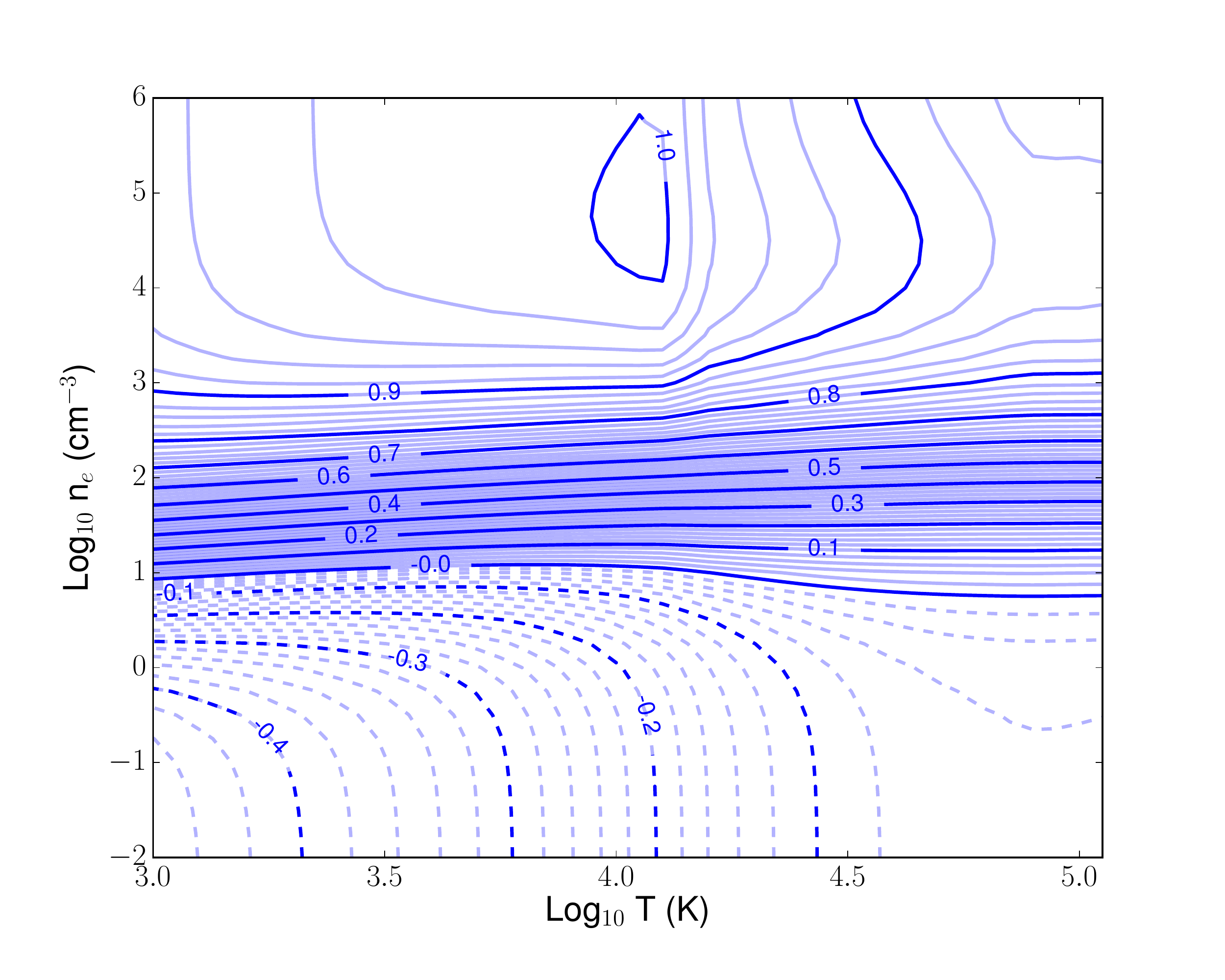}
\includegraphics[width=85mm]{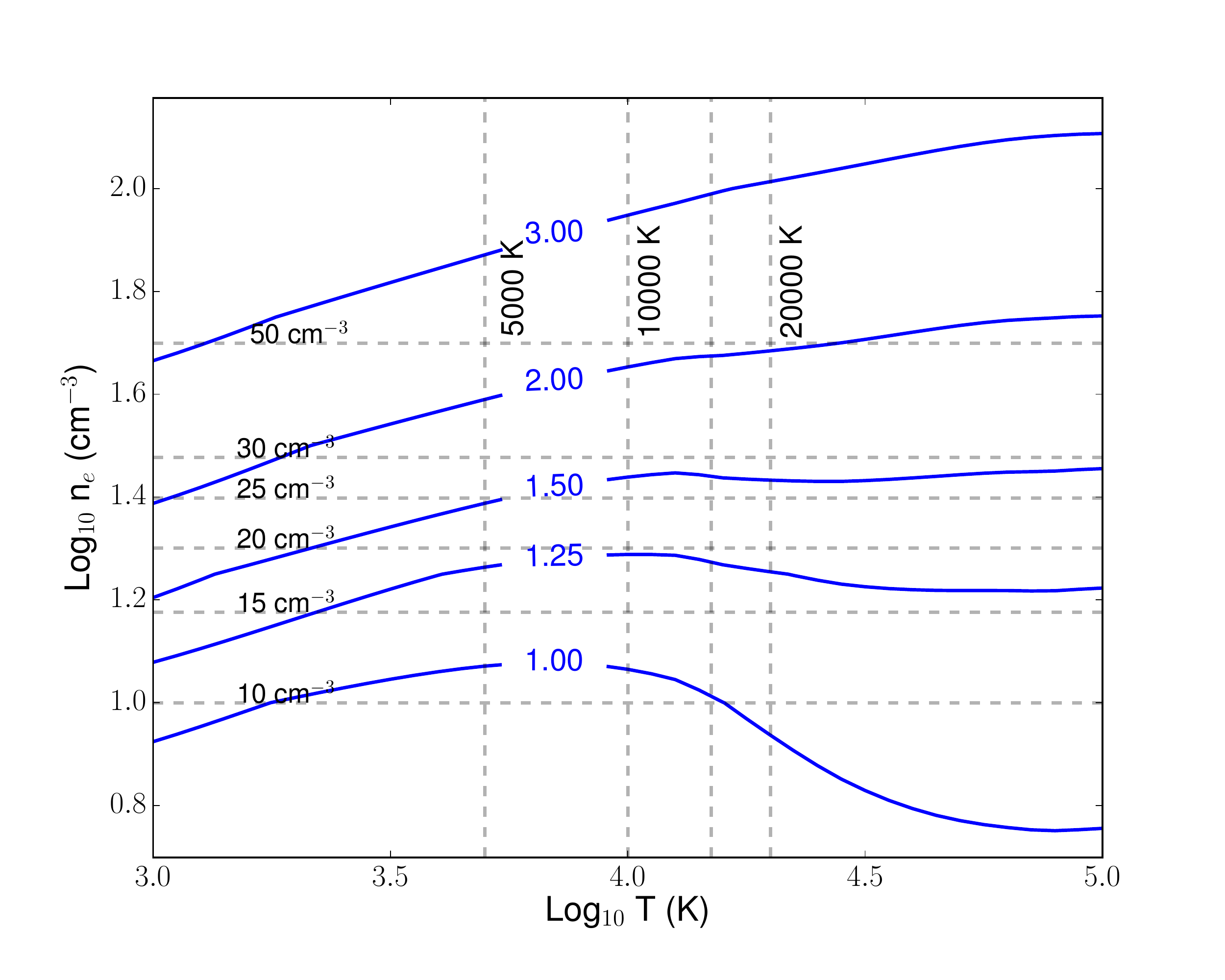}
\vskip-0.1truecm
\caption{Log$_{10}$ of the line ratio $I({\rm N^+}\;122\,{\micron})/I({\rm N^+}\;205\,{\micron})$, over a broad range of temperature and electron density (top panel). Unlabelled contours are spaced at intervals of 0.02 in the logarithm. Most of the line ratios measured by G15 fall in the range $1-3$, for which the inferred value of $n_e$ is seen to be insensitive to the temperature. Bottom panel: detail of the intensity ratios that fall in the range 1 to 3.}
\label{fig:N122N205}
\end{figure}

\section{Constraints from Herschel Data}

\subsection{Is high electron density required?}
G15 presented good reasons to expect that their derived electron densities are robust. However, as noted in the introduction, the problems of interpretation posed by the D-WIM are challenging, so we sought to verify the G15 analysis of the N$^+$ line ratio, using \chianti. We therefore computed the expected intensity ratio of N$^+\;122\,{\rm\mu m}$ to N$^+\;205\,{\rm\mu m}$, over a wide range of density and temperature, with the results as shown in figure \ref{fig:N122N205}. Our ratios coincide closely with those of G15, at their assumed temperature of  $T=8{,}000\,{\rm K}$, and we confirm their conclusions that modest temperature variations around that value yield only minor variations in the derived electron density. Furthermore, most of the line ratios measured by G15 fall in the range 1 to 3, for which the derived electron densities are insensitive to the assumed temperature over a very broad range indeed. We conclude that the G15 estimates of electron density are robust to changes in the assumed temperature.

\subsection{C$^+\,158\,{\rm\mu m}$ emission}
\begin{figure}
\includegraphics[width=85mm]{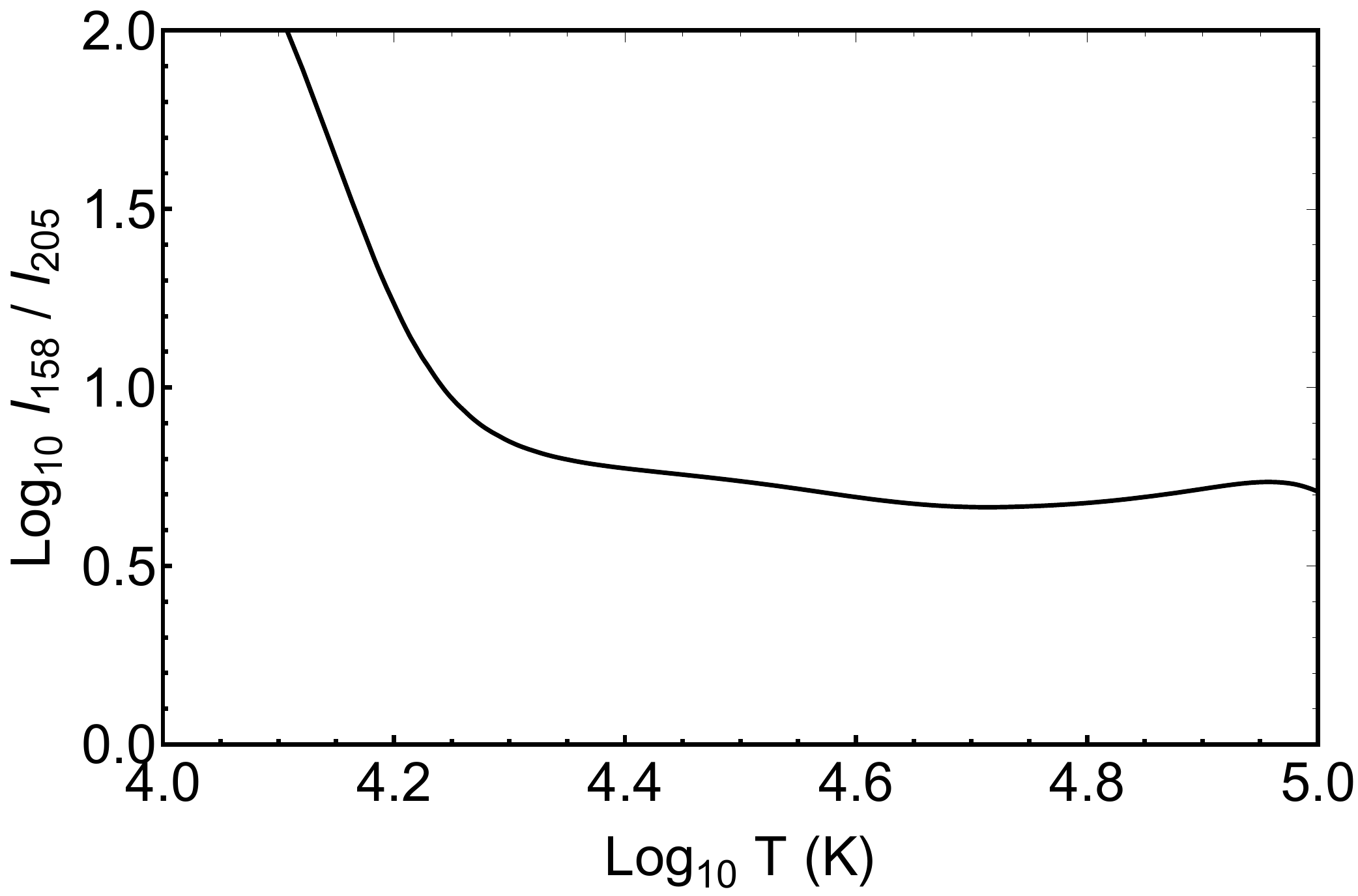}
\vskip-0.1truecm
\caption{Calculated ratio of C$^+\;(158\,{\rm \mu m})$ to N$^+\;(205\,{\rm \mu m})$ line intensities, as a function of temperature, for our model plasma with $n_e=30\,{\rm cm^{-3}}$ (\S2.2).}
\label{fig:C158N205}
\end{figure}

Emission from C$^+\;(158\,{\rm\mu m})$ is also seen in the inner Galactic plane, and correlates strongly with the intensity of the nitrogen lines (G15). That correlation may arise because of emission from the dense plasma itself, or possibly from neutral gas that is in some way associated with the dense plasma.\footnote{Although the $158\,{\rm\mu m}$ line is from an ion, it may exist in gas which is largely neutral because the first ionisation potential of carbon is lower than that of hydrogen.} An acceptable plasma model must not produce more C$^+$ emission than is actually observed. In their table 4, G15 reported C$^+\;(158\,{\rm\mu m})$ and N$^+\;(205\,{\rm\mu m})$ intensities on ten lines-of-sight in the inner Galactic plane; their mean line ratio is 7.4, with a standard deviation of 1.5. Our predicted ratio of C$^+\;(158\,{\rm\mu m})$ to N$^+\;(205\,{\rm\mu m})$ line intensity is shown in figure \ref{fig:C158N205}, as a function of temperature. From it, we deduce a plasma temperature of $T=19{,}500\,(+5{,}600,-1{,}400)\,{\rm K}$ to match the observed mean line ratio minus or plus one standard deviation.

The very steep temperature dependence of the line ratio, below $T=19{,}500\,{\rm K}$, means that even slightly lower temperatures yield too much C$^+$ emission in the inner Galaxy. For example: at $T=15{,}500\,{\rm K}$ the predicted line ratio is approximately 20, which is too large.

We caution that the foregoing analysis relies on the implicit assumption that the observed C$^+$ and N$^+$ FIR lines are both optically thin, which might not be the case in practice. Indeed \citet{langer2016} argued that significant optical depth of C$^+\;(158\,{\rm \mu m})$ is sometimes indicated by a comparison of the C$^+$ and N$^+$ line profiles. However, the strong and roughly linear correlation that is observed between C$^+$ and N$^+$ line strengths (G15) suggests that high optical depth is not typical.

We consider $T\simeq19{,}500\,{\rm K}$ to be a preferred temperature for our model plasma, because it supplies a simple explanation for the origin of the observed (G15) strong correlation between C$^+$ and N$^+$ FIR line emissions. In this picture, ionised gas is the dominant source of C$^+\;(158\,{\rm\mu m})$ emission in the inner Galaxy --- an idea originally suggested by \citet{heiles1994}. Our inference thus differs from the analysis of G15, who concluded that only 30-50\% of the C$^+$ emission arises from dense plasma. In part this difference is due to the larger carbon-to-nitrogen abundance ratio in our model, which is $4.0$ versus $2.9$ in G15. The other contributing factor is that G15 implicitly assumed the nitrogen to be entirely in the form N$^+$, whereas our plasma model uses the ionised fraction that corresponds to collisional ionisation equilibrium at the specified temperature and density. 

In \S4.2 we will make a second estimate of the temperature of the D-WIM ($18{,}400\,{\rm K}$), by comparing N$^+$ emission to microwave bremsstrahlung. We have no reason to prefer that estimate over the current one, or vice versa, so we take the approximate midpoint, $T=19{,}000\,{\rm K}$, as our best estimate of the D-WIM temperature in the inner Galactic plane. With that temperature, and $n_e=30\,{\rm cm^{-3}}$, we can convert N$^+\;(122\,{\rm\mu m})$ line intensities to emission measures using the relationship ${\rm EM}({\rm pc\, cm^{-6}})=\epsilon\,I_{122}({\rm W\,m^{-2}\,sr^{-1}})$, with $\epsilon=1.74\times10^{11}$. 

\section{Constraints from Planck data}
The {\it COBE\/} FIRAS data \citep{bennett1994} demonstrated a disk morphology for the Galactic N$^+\;{\rm(205\,\mu m)}$ FIR line emission, with a strong concentration towards the inner Galaxy. However, the scale height of the disk was not well resolved by the $7^\circ$ beam of the {\it FIRAS\/} instrument. Much better angular resolution is available in the bremsstrahlung (free-free) continuum map generated from the \planck\ all-sky microwave data. As the ionisation potential of nitrogen is greater than that of hydrogen, the N$^+$ line emission must arise in the ionised medium, so there is a strong expectation that the bremsstrahlung continuum and the N$^+$ line emission should trace the same gas.

We have tested that expectation by correlating the \planck\ free-free map with the N$^+$ FIR data in the inner Galaxy, where the latter signal is detected. For each dataset, we tried  different quantities that gauge the amount of ionised gas on the line-of-sight. For the \herschel\ data we used four quantities, which we label H1 to H4, as follows: H1, the N$^+\;122\,{\rm\mu m}$ intensity; H2, the N$^+\;205\,{\rm\mu m}$ intensity; H3, the N$^+$ column; and H4, the product of that column with the electron density (i.e. a proxy for the EM). In each case we used only those lines-of-sight for which G15 obtained detections of both N$^+\;122\,{\rm\mu m}$ and N$^+\;205\,{\rm\mu m}$, so that all four measures are available. For the \planck\ data we used two quantities, which we label P1 and P2, as follows: P1, the EM as given in the \planck\ foreground map; and P2, the 30~GHz free-free continuum intensity, which we computed from the EM and temperature, as given in the \planck\ map, using the formulae given in table 4 of \citet{adam2016}. The \planck\ free-free map actually quotes two values for both EM and temperature, for every pixel: one is the mean value of the posterior probability distribution that is obtained from the multi-frequency modelling, and the other is the mode of that probability distribution. The two are very similar where the signals are strong, as is the case in the Galactic plane, but the mean is more useful at high latitudes where the signals are weak and the mode is often zero. Throughout this paper, therefore, we use the mean of the posterior distributions for all of the \planck\ data products. In this way we obtained 8 correlation coefficients for the set of 92 locations where G15 detected both N$^+$ FIR lines.

\begin{figure}
\includegraphics[width=85mm]{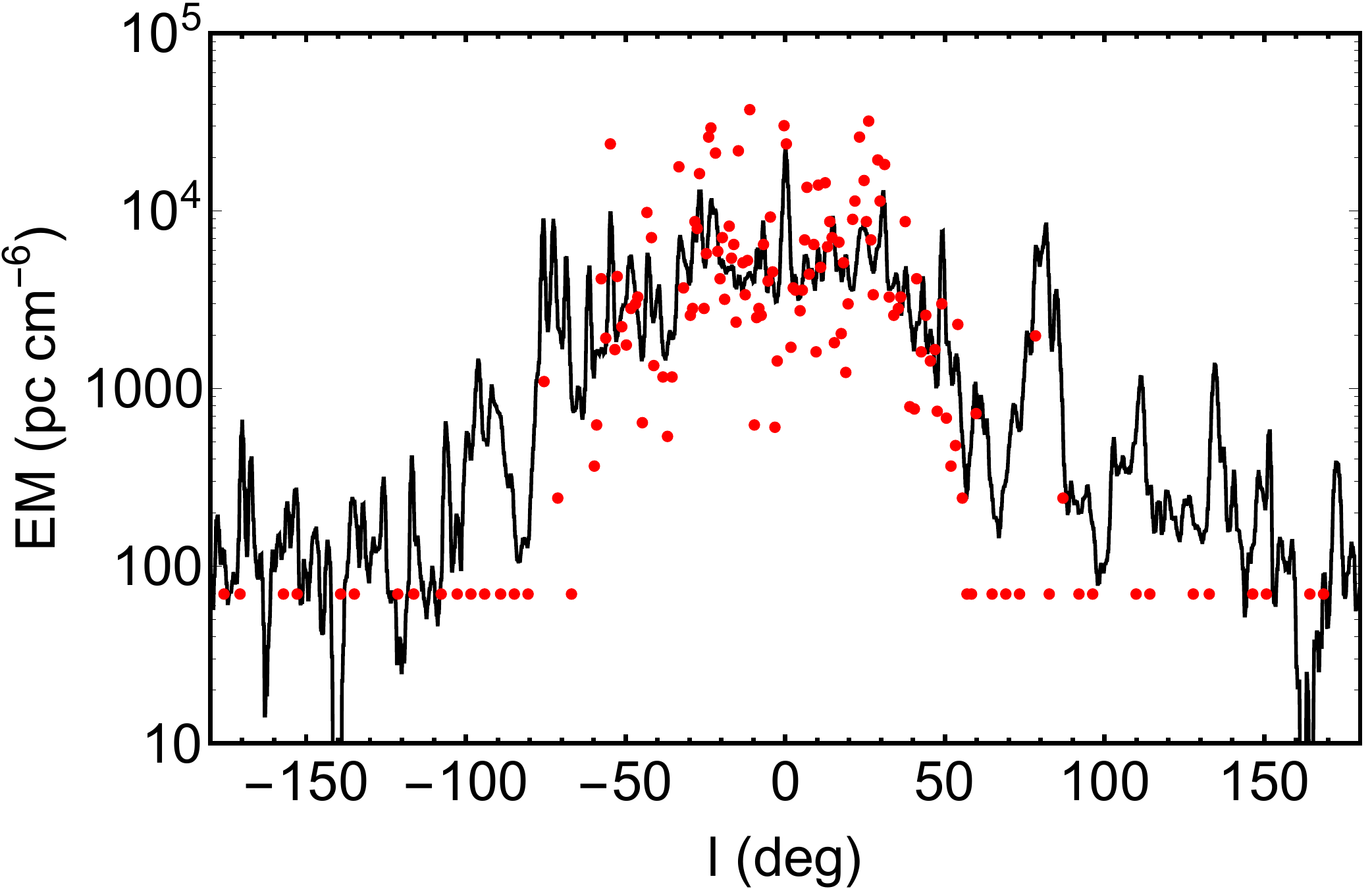}
\vskip-0.1truecm
\caption{The Galactic longitude variation of emission measure, as determined from \herschel\ data (N$^+\;122\,{\rm \mu m}$; red points), and as given in the \planck\ free-free foreground map (black line). All data are for the Galactic plane, but the effective \planck\ resolution is $\sim1^\circ$, whereas the \herschel\ PACS field-of-view is $\sim1^{\prime}$.}
\label{fig:EM_GalL_HerschelPlanck}
\end{figure}

We found that the four correlations taken with P1 were very similar to their counterparts taken with P2 -- fractional differences between the two range from 0.1\% to 0.7\% -- and in 2/4 cases P1 correlated better than P2. Thus we take P1 and P2 to be equivalent measures, and we only report the mean of the correlations with P1 and P2 in each case. The best correlation was with H2  (0.73), followed by H3 (0.69), H1 (0.67) and H4 (0.64). We do not have a ready explanation for why N$^+\;205\,{\rm\mu m}$ yields the highest correlation. For our purposes the main point is that there is indeed a high correlation between the FIR and microwave tracers of the ionised gas. Because of the lower angular resolution of the \planck\ data (of order one degree), compared to  \herschel\ PACS (of order one arcminute, after averaging over the field-of-view), we do not expect a perfect correlation between the two. We therefore regard the observed high correlation as a confirmation that the two signals have a common origin, i.e. in the inner Galactic plane both are dominated by the D-WIM.

Figure \ref{fig:EM_GalL_HerschelPlanck} shows the longitude profile of EM in the \planck\ free-free map, along with the EM inferred from the N$^+$ $122\,{\rm \mu m}$ emission reported by G15, using the conversion factor given in \S3.2. Where G15 reported an upper limit to the line intensity we have adopted a common, representative value, yielding the plateau of red points at ${\rm EM}\simeq70\,{\rm pc\,cm^{-6}}$. In the inner Galaxy, where N$^+$ $122\,{\rm \mu m}$ is detected, the EM inferred from the G15 data shows much stronger point-to-point variations than the \planck\ estimate of EM. That is not surprising, given the lower angular resolution of the latter. It also appears that the mean of the G15 estimates is systematically higher than for \planck. That suggestion is borne out quantitatively, as follows. For the same set of data used in the correlation analysis, we have computed the mean of the EMs inferred from the N$^+$ $122\,{\rm \mu m}$ intensity, and the mean of the EMs given in the \planck\ free-free map. We find that the ratio of these mean values is 1.76. Part of this difference can be understood as a consequence of the lower angular resolution of \planck, as described in the next section.

\begin{figure}
\includegraphics[width=85mm]{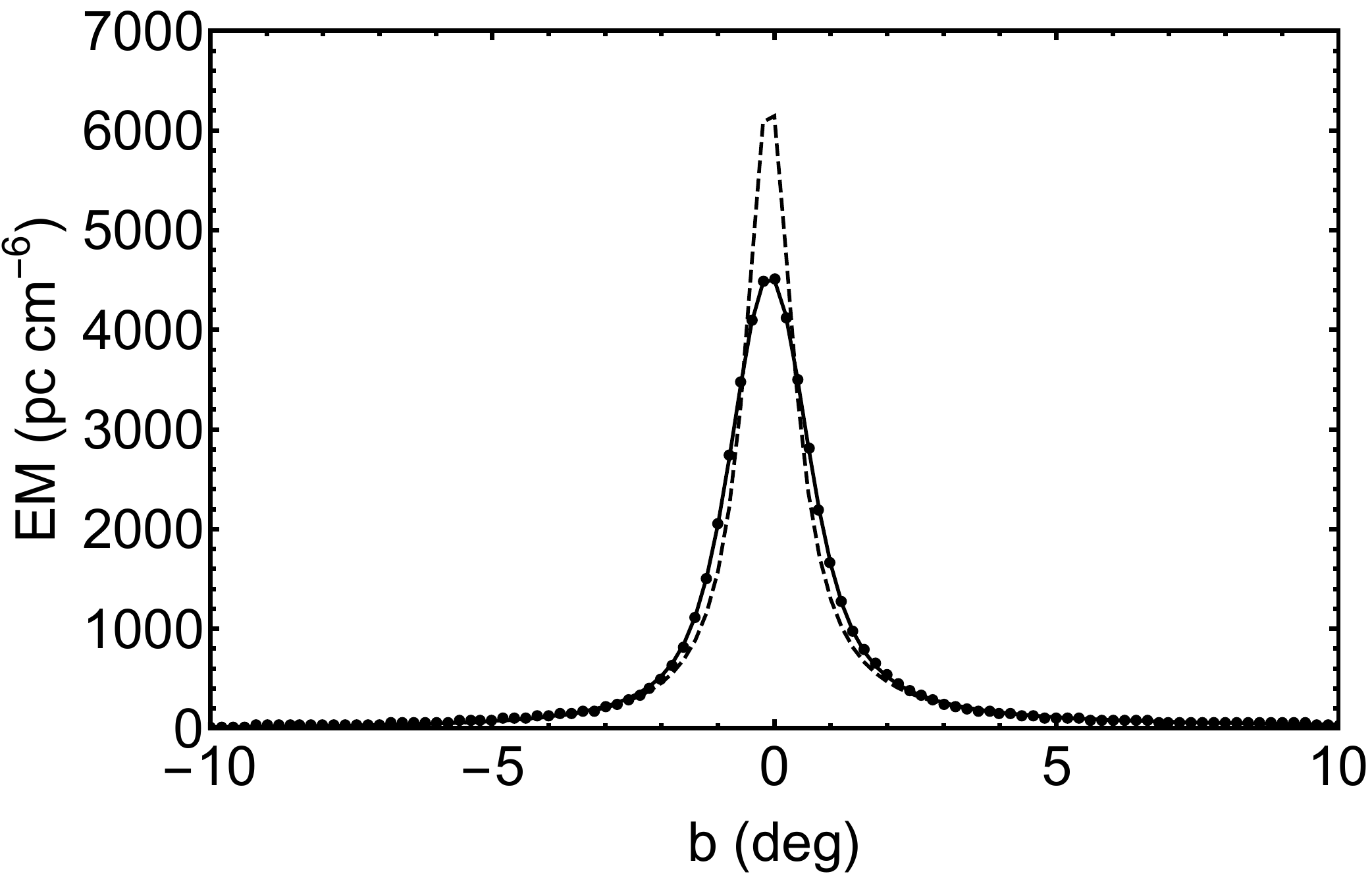}
\vskip-0.1truecm
\caption{The Galactic latitude dependence of EM attributed to the free-free process in the \planck\ foreground model of \citet{adam2016}. The \planck\ data are shown as black points, and have been averaged over the region $|l|\le60^\circ$. The dashed line shows our model, equation  (1). The solid line shows the convolution of that model with a Gaussian profile (${\rm FWHM}=1.07^\circ$), representing the point-spread function of the \planck\ map.}
\label{fig:EM_GalB_Planck}
\end{figure}

\subsection{Disk thickness}
We now use the \planck\ data to determine the thickness of the disk of D-WIM. Figure \ref{fig:EM_GalB_Planck} shows the longitude-averaged variation in EM, as a function of Galactic latitude, $b$, in the \planck\ free-free map of \citet{adam2016}. The point-spread-function (PSF) of the map is expected to be approximately Gaussian with full-width at half maximum (FWHM) equal to $1.07^\circ$ --- slightly larger than the notional FWHM of $1^\circ$, as a result of gridding. At very small values of $|b|$ the data exhibit a form that is similar to the PSF, but the peak is noticeably offset from zero, being centred on $-0.1^\circ$. There is also an asymmetry in the wings of the profile, in the opposite sense with the signal in the range $5^\circ\la b\la10^\circ$ being higher than that in the range $-5^\circ\ga b\ga-10^\circ$. Both wings lie well above the PSF, but they decline rapidly with increasing $|b|$. (Figure 6 shows the same data in a logarithmic scale, where the wings are seen more clearly.) 

We have modelled the EM distribution shown in figure~\ref{fig:EM_GalB_Planck} using a power-law with an inner scale:
\be
{\rm EM}(b)={{{\rm EM}(0)}\over{(1+(b/b_o)^2)^{\alpha/2}}},
\ee
which is then offset by $-0.1^\circ$, and tilted slightly (a multiplicative factor of $1+0.049 b$) to accommodate the asymmetries in the observed distribution. By convolving this distribution with the Gaussian PSF and matching to the data, we obtain the fit shown in figure \ref{fig:EM_GalB_Planck} with the parameters $b_o=0.488^\circ$, $\alpha=1.82$ and ${\rm EM}(0)=0.638\times10^4\,{\rm pc\,cm^{-6}}$. At an expected median distance of $D=8.2\,{\rm kpc}$, our inferred inner scale, $b_o$, corresponds to a height of $z_o=b_oD\simeq70\,{\rm pc}$, and the FWHM of the profile (equation 1) is then $130\,{\rm pc}$.

The linear scale in figure 4 gives prominence to the bright core of the intensity distribution, at the expense of the fainter wings. Subsequently we will render the same data on a logarithmic scale (figure 6), which reveals the wings more clearly. There is no reason to suppose that the wings of the profile have a fundamentally different origin to the core -- on the contrary, our power-law distribution yields a good model -- so in addition to the bright, thin disk, we conclude that faint emission from the D-WIM is observable out to beyond $1\,{\rm kpc}$ from the plane. Because of its greater scale height this material gains in prominence, relative to the thin disk, when we observe at high latitudes, and it thus comes to the fore when we consider the local manifestations of D-WIM (\S6).

The vertical distribution that we have determined for the D-WIM is very similar to that reported for the C$^+$ emission in the inner Galaxy \citep{velusamy2014}. Both profiles exhibit a strong, narrow peak, offset slightly to the South of the plane, rolling over into extended wings at higher latitudes. This suggests that the C$^+$ emission is associated with the D-WIM. On the other hand \citet{velusamy2014} associated the strong, narrow peak of the C$^+$ intensity with CO-emitting molecular gas (FWHM 129$\,{\rm pc}$), and the broader wings of the profile with diffuse \htwo\ and the WIM. Taken together the associations of C$^+$ with both CO and D-WIM imply that the D-WIM is somehow associated with CO-emitting molecular gas. Recall that in \S3.2 we argued on the basis of the C$^+$/N$^+$ line ratios that the C$^+$ emission arises primarily from the D-WIM itself \citep[see also][]{abel2006}, but the nature of the association between the D-WIM and the CO-emitting molecular gas remains unclear.

\begin{figure}
\includegraphics[width=85mm]{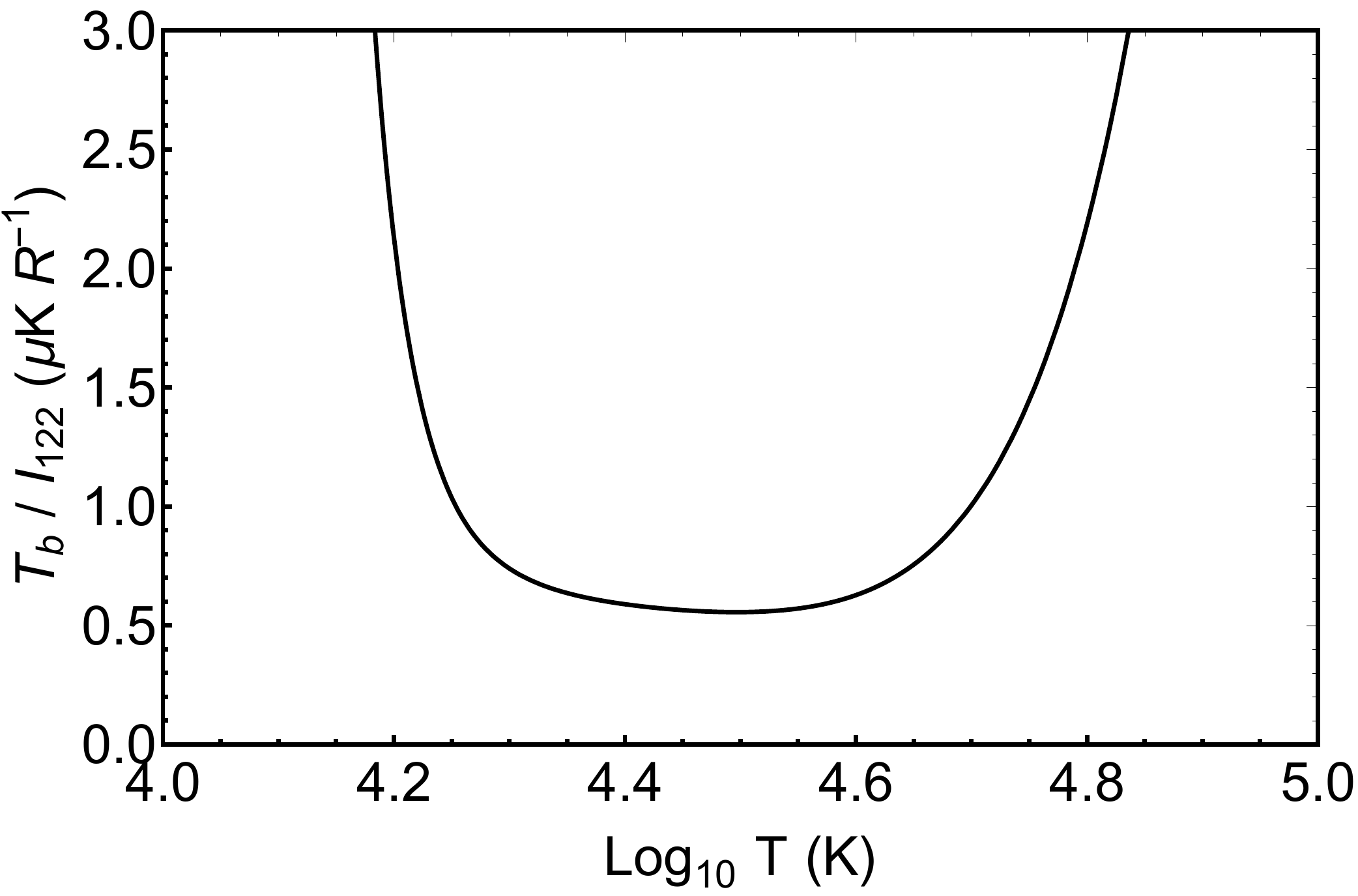}
\vskip-0.1truecm
\caption{The predicted ratio of bremsstrahlung intensity at $\nu=22.8\,{\rm GHz}$ to the N$^+\;(122\,{\rm\mu m})$ line intensity, as a function of temperature, for our plasma model with $n_e=30\,{\rm cm^{-3}}$.}
\label{fig:microw_N122}
\end{figure}

\subsection{Temperature of the plasma}
As is evident in figure \ref{fig:EM_GalB_Planck}, the value of EM at $b=0$ in the \planck\ map is significantly lower than the underlying value, because \planck\ has not fully resolved the vertical structure of the emission. The ratio of our model ${\rm EM}(0)$ to the observed value is a factor of 1.365. This explains part, but not all of the systematic offset between \planck\ EMs and \herschel\ EMs that was mentioned in the previous section. The remaining discrepancy is easily understood: the EMs we inferred from the N$^+$ $122\,{\rm \mu m}$ line use our preferred temperature of $19{,}000\,{\rm K}$ (\S3.2), whereas the EMs in the \planck\ free-free map correspond to the temperatures that are also specified, pixel-by-pixel, in the \planck\ free-free map. Those temperatures are much lower than our preferred value. Now the microwave brightness temperature due to bremsstrahlung declines with increasing plasma temperature, roughly as the square-root, so for a given free-free intensity determined by \planck, a higher plasma temperature corresponds to a larger EM.

There is in fact very little information on the plasma temperature in the \planck\ data themselves: the bremsstrahlung intensity is strongly constrained, but the plasma temperature and the emission measure are almost completely degenerate with each other in the fitting process. We are therefore at liberty to interpret the \planck\ bremsstrahlung map in terms of emission from a hotter plasma, with a correspondingly higher EM.  As the N$^+$ line and microwave continuum intensities both scale linearly with EM, the ratio of the two can be used to determine the plasma temperature in the context of our model.

Taking the same lines-of-sight that we used in our correlation analysis, we find that the mean \herschel\ N$^+$ $122\,{\rm \mu m}$ intensity\footnote{The unit is Rayleigh: $1\,{\rm R=10^6/(4\pi)\,photons\,cm^{-2}\,s^{-1}\,sr^{-1}}$} is $4.25\times10^4\,{\rm R}$. The mean \planck\ free-free brightness temperature at\footnote{We use $\nu=22.8\,{\rm GHz}$ for consistency with previous studies of the microwave continuum strength, relative to \ha\ --- see \S6.1.} $\nu=22.8\,{\rm GHz}$, for the same directions, is $T_b=2.84\times10^4\,{\rm \mu K}$. Correcting for beam smearing in the \planck\ map involves a factor of 1.365, as described above, leading to an estimate of $0.91\,{\rm \mu K\,R^{-1}}$ for the microwave-to-FIR ratio. Figure \ref{fig:microw_N122} shows the ratio that is expected for our plasma model, at $n_e=30\,{\rm cm^{-3}}$, as a function of temperature. To match the data on the inner Galaxy requires either a temperature $T\simeq18{,}400\,{\rm K}$, or else $T\simeq48{,}000\,{\rm K}$. Of these two possible solutions we prefer the former as it is consistent with our temperature determination based on the C$^+$/N$^+$ ratio (\S3.2). (We also note that the optical emission line ratios measured at high latitude provide strong evidence against temperatures as high as $48{,}000\,{\rm K}$ for the local WIM --- see \S6, and particularly figure 12.)

With the plasma temperature determined in this way, the emission measures implied for the \planck\ bremsstrahlung foreground are larger than the EM values given in the map of \citet{adam2016}, by a factor of approximately 1.3. Throughout the rest of this paper, unless explicitly stated otherwise, the EM values that we determine from the \planck\ free-free map are those appropriate to $T=19{,}000\,{\rm K}$ --- the approximate midpoint of the two values we have determined for the D-WIM in the inner Galaxy, from the ratio of bremsstrahlung to N$^+$ line intensity, and from C$^+$ to N$^+$ line intensity (\S3.2).

\subsection{Small scale structure}
We have already noted that there is more point-to-point fluctuation in the N$^+$ intensity measurements than in the microwave data. Specifically, in the inner Galaxy ($|l|\la60^\circ$), the red points in figure \ref{fig:EM_GalL_HerschelPlanck} exhibit both higher peaks and deeper troughs than the black line. Both datasets have high signal-to-noise ratio in this region, so these differences are certainly attributes of the signals sampled by the two instruments. As we have already established that both are dominated by the D-WIM, the most likely explanation for the different fluctuation levels is the finer angular resolution of the G15 intensity measurements compared to the \planck\ free-free map: $47^{\prime\prime}$ for the former, and $\simeq1^\circ$ for the latter.  In this interpretation, the D-WIM is highly structured on an angular scale that is small compared to the \planck\ resolution. 

We note that the images presented in G15's figure 11 all show significant structure on scales smaller than the PACS field of view, corresponding to length scales $\la2\,{\rm pc}$ at a median distance of 8.2~kpc.

\section{Contribution to pulse dispersion}
Pulse dispersions are precisely measured for almost all of the $\sim2{,}500$ known radio pulsars in the Galaxy.\footnote{http://www.atnf.csiro.au/research/pulsar/psrcat} However, as the pulsars are not at infinite distance, the resulting information on DMs can only be related to EMs -- which sample all the plasma on the line-of-sight -- via the construct of a three-dimensional model. Unfortunately only a small fraction of pulsars are useful in that endeavour, as most lack  reliable distance measurements. With only sparse spatial sampling of the Galaxy, the resulting electron density models are necessarily simplified, smooth descriptions of a medium which is likely to be highly structured. In this paper we rely on published models of the free electron distribution, based on pulsar DMs.

At the time of writing there are three such models that are commonly in use: TC93 \citep{taylorcordes1993}; NE2001 \citep{cordeslazio2002}; and YMW16 \citep{yao2017}. Inevitably there is much in common across these three models. In particular, all three contain an axisymmetric thin disk, an axisymmetric thick disk, and spiral arm components. However, although these features have similar properties in TC93 and NE2001, their appearance is very different in YMW16, who follow an updated model presented in \citet{houhan2014}. A significant difference between YMW16 and the preceding models, is that YMW16 does not include scattering measures in its construction, nor does it apply voids or clumps towards particular pulsars with discrepant DM values. The motivation is that scattering measures are often dominated by a few regions of strong $n_e$ fluctuations along the line of sight \citep[e.g.][]{stinebring2001,brisken2010}, and several studies have consequently revealed deviations from the established scattering measure vs. DM trends (e.g. \citealt{lewandowski2015, geyer2017}).

In this paper we make use of both TC93 and YMW16 to demonstrate the range of possible representations of the major structural features of the Galactic free-electron distribution.

In the case of TC93 we used the functional forms and parameters for the various components of the $n_e$ distribution, as stated by TC93, and integrated along the line-of-sight to determine the DM. For YMW16, we took the code provided by the authors and modified it appropriately to determine the contribution of each of the components of interest.\footnote{We note that the original YMW16 code combines the contribution of the spiral arms and the thin disk using a {\tt max()} function, so that the YMW16 model is not simply a sum of its individual components.}

\begin{figure}
\includegraphics[width=85mm]{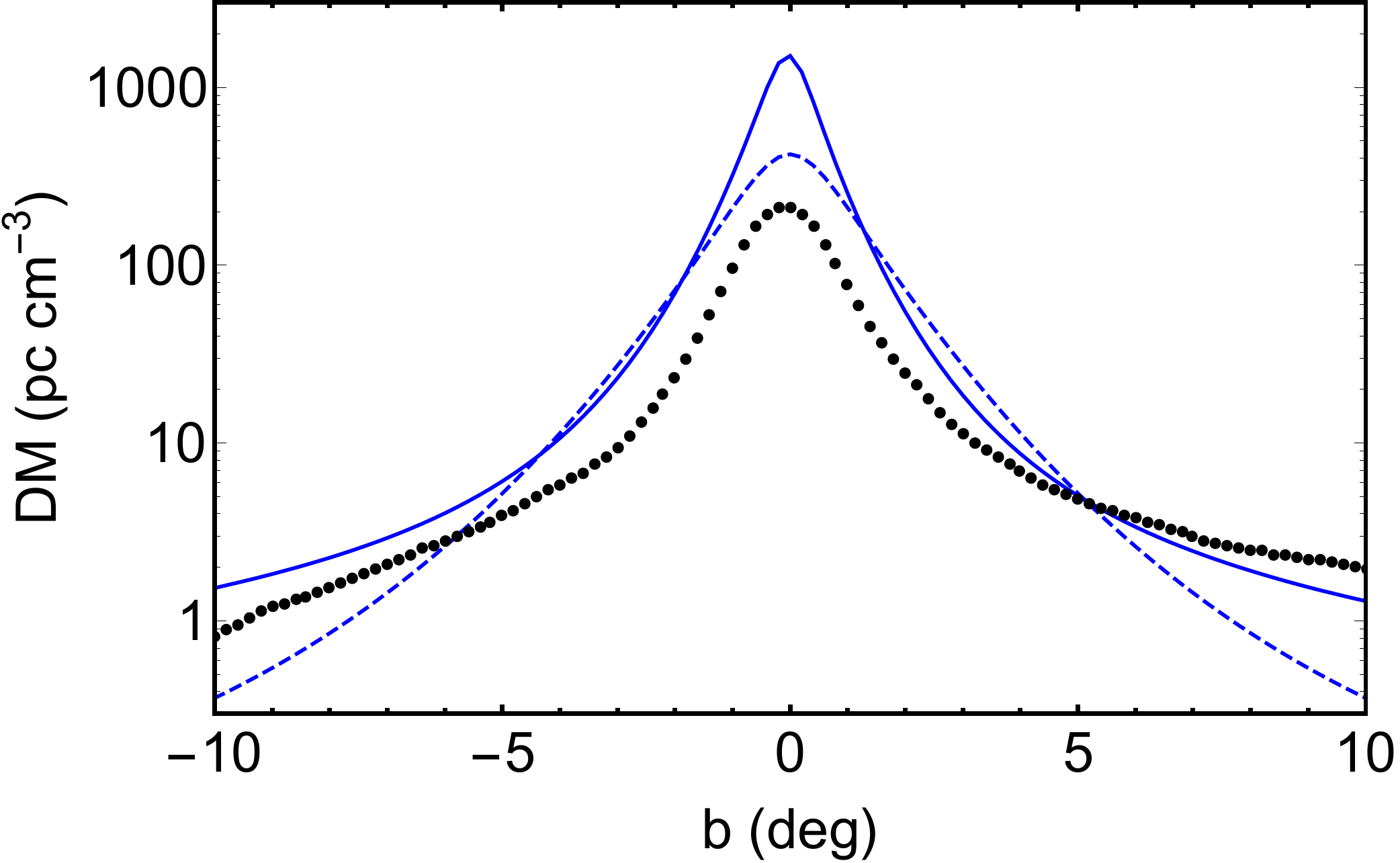}
\vskip-0.1truecm
\caption{DM contribution of the D-WIM as a function of Galactic latitude, averaged over $|l|\le 60^\circ$. Black points are the \planck\ bremsstrahlung component as per figure \ref{fig:EM_GalB_Planck}, but now with EMs computed for $T=19{,}000\,{\rm K}$, and rescaled according to ${\rm DM}={\rm EM}/n_g$. Also shown are the thin disk components of the TC93 (dashed blue curve) and YMW16 (solid blue curve) free electron models.}
\label{fig:DM_GalB_Planck_thindisk}
\end{figure}

\subsection{Thick disk}
Towards the Galactic poles, the thick disk component contributes $17.1\,{\rm pc\,cm^{-3}}$  in YMW16 (average of North and South), and $16.5\,{\rm pc\,cm^{-3}}$ in TC93. It is easy to see that these DMs cannot arise in dense gas, with $n_e\sim30\,{\rm cm^{-3}}$, because the implied emission measure would then be ${\rm EM}\sim500\,{\rm pc\,cm^{-6}}$, whereas the observed values are $\sim1\,{\rm pc\,cm^{-6}}$ (see \S6). We conclude that, for either of the free-electron models, the thick disk component cannot be made of \mbox{D-WIM}. In both free-electron models the thick disk is the dominant contribution to the DM towards the poles, so this can be stated in a model independent way: the gas which contributes the majority of the DM at high latitudes is not the D-WIM.

\begin{figure*}
\includegraphics[width=85mm]{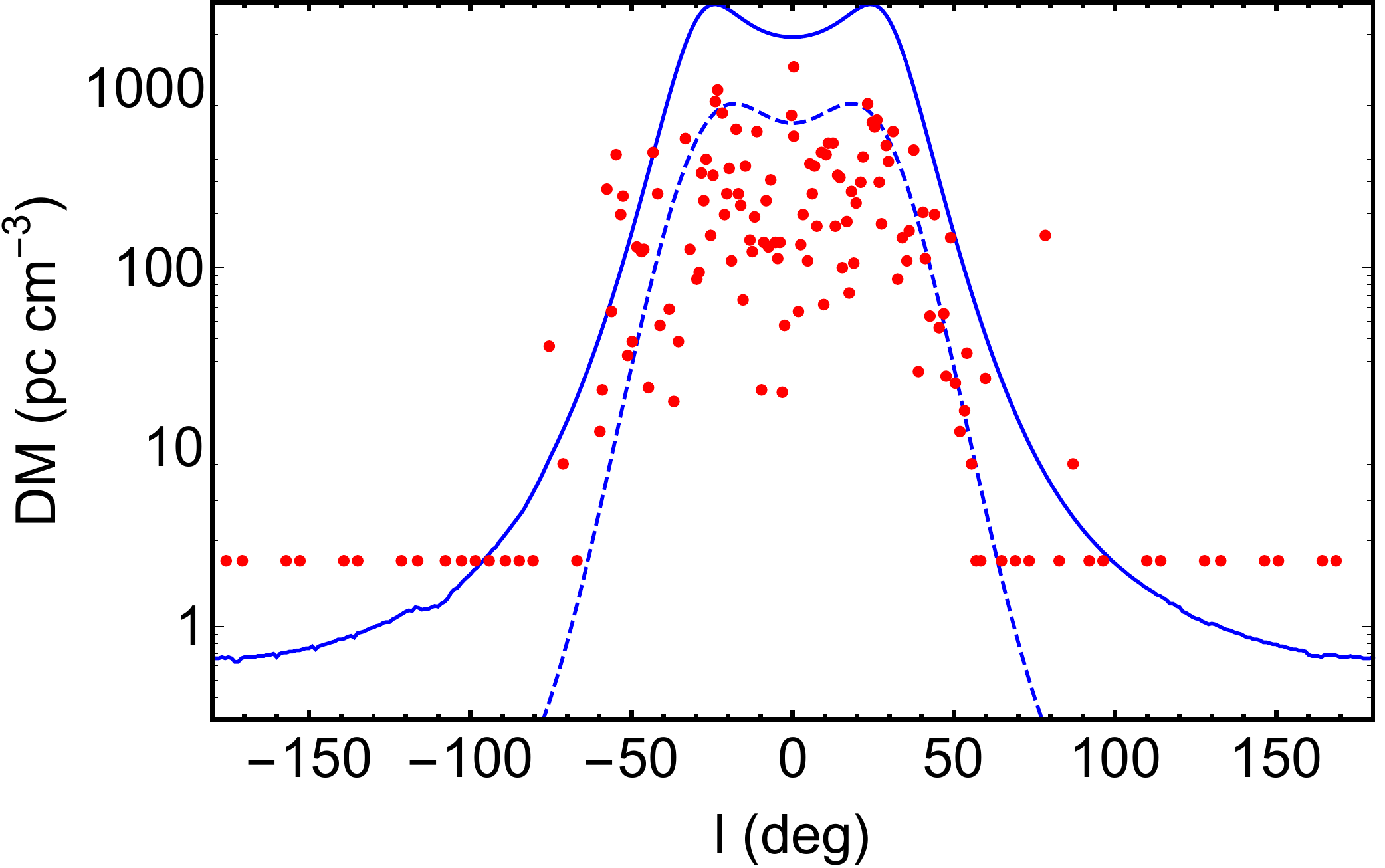}\ \ \includegraphics[width=85mm]{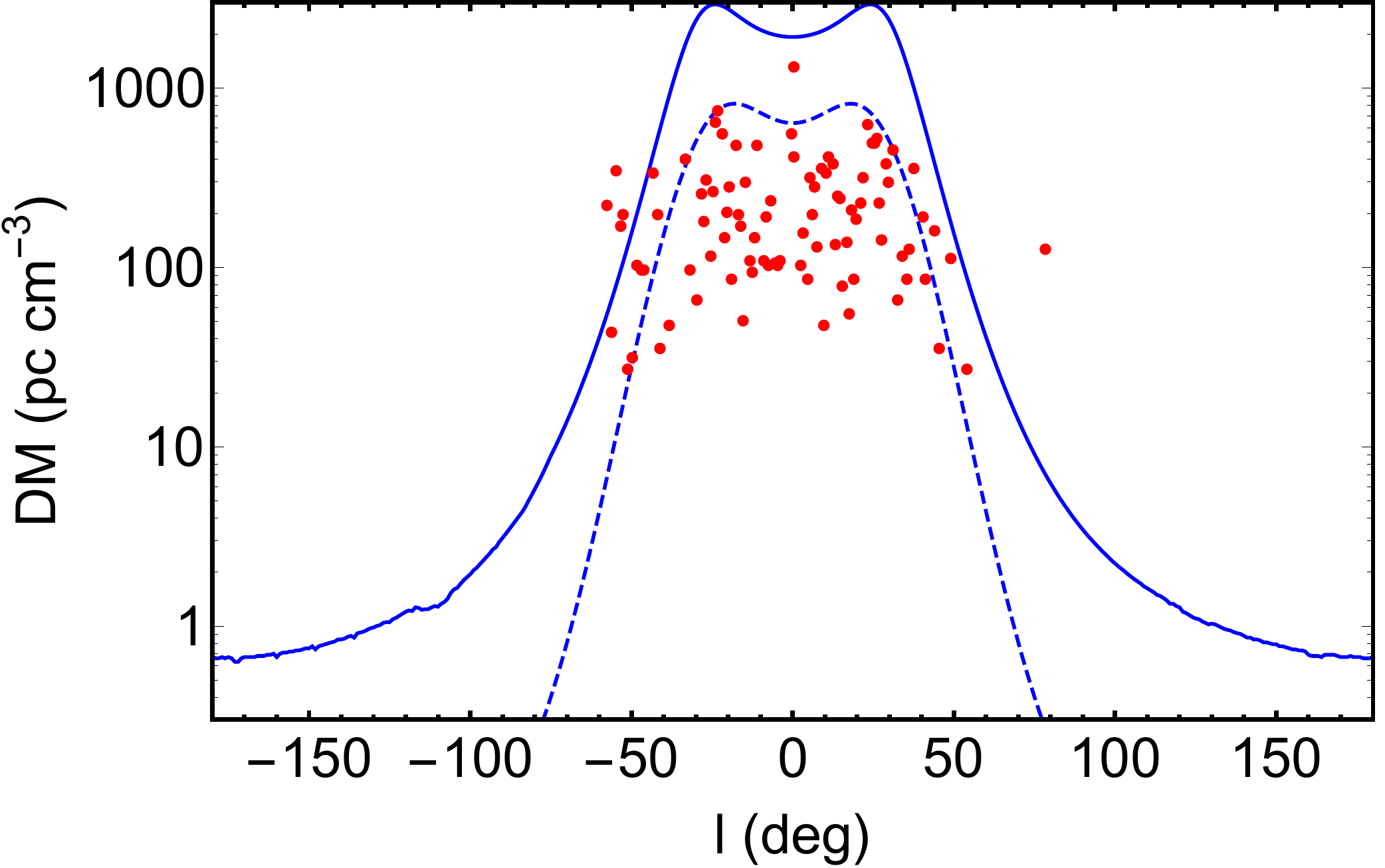}
\vskip-0.1truecm
\caption{Inferred and model DMs as a function of Galactic longitude. Left panel: the red points show estimates of the DM contribution of the dense plasma responsible for the N$^+\;(122\,{\rm \mu m})$ emission, using the line intensity reported by G15, the intensity to EM conversion given in \S3.2, and rescaled according to ${\rm DM}={\rm EM}/n_e$. For lines of sight where N$^+\;(205\,{\rm \mu m})$ was not detected (so $n_e$ could not be determined) we have adopted $n_e=n_g=30\,{\rm cm^{-3}}$. Right panel: the red points show lower limits on DM evaluated from the column-density of N$^+$ ions determined by G15.  In both panels we show the thin disk component of the TC93 (dashed blue curves) and YMW16 (solid blue curves) free electron models.}
\label{fig:DM_GalL_Herschel}
\end{figure*}

\subsection{Galactic Centre}
We note that the largest DM value inferred from \herschel\ data happens to coincide with the Galactic Centre (GC). In the YMW16 and NE2001 model distributions, an additional component of free electrons is located there, leading to a spike in the predicted DM for sources in or behind the GC. The effect can be seen, for example, in figure 1 of YMW16. Indeed the DM spike expected on the basis of the \herschel\ data is $1{,}293\,{\rm pc\,cm^{-3}}$, which is similar in amplitude to the GC contribution seen in YMW16's figure 1. We conclude, therefore, that the GC DM component in the free electron models is attributable to the D-WIM reported by G15.

\subsection{Thin disk}
We have already established that the dense plasma discovered by G15 takes the form of a thin disk, concentrated in the inner Galaxy. Our inferred FWHM of approximately $130\,{\rm pc}$, determined in \S4.1, is similar to that of the thin disk component in YMW16, and smaller than that of TC93 (approximately $260\,{\rm pc}$). As described in \S2, for the D-WIM we can readily convert between EM and DM because we know the characteristic electron density in the emitting regions. Accordingly, figure \ref{fig:DM_GalB_Planck_thindisk} shows the \planck\ bremsstrahlung component as per figure \ref{fig:EM_GalB_Planck}, but now with EMs evaluated at $T=19{,}000\,{\rm K}$ and then scaled by $1/n_g$ to yield the DM. The YMW16 and TC93 thin disk components are also shown. The TC93 structure is broader than the \planck\ profile, but its normalisation is only a factor $\sim2$ larger; for YMW16, on the other hand, the angular width of the disk is similar to that seen with \planck, but the normalisation is an order of magnitude larger. In both models it is the thin disk component of free electrons that is best matched to the observed latitude profile of the D-WIM, as we will see.

The small scale-height and concentration towards the inner disk of the Galaxy both lead us to expect that the thin disk should make only a minor contribution to the DM observed towards the poles. That is indeed the case in practice, with YMW16 predicting a contribution of $0.14\,{\rm pc\,cm^{-3}}$ (average of North and South), and TC93 predicting $7\times10^{-3}\,{\rm pc\,cm^{-3}}$; the corresponding EM values are $4\,{\rm pc\,cm^{-6}}$ and $0.2\,{\rm pc\,cm^{-6}}$, respectively. The EM required to explain the high-latitude emission from the WIM lies between these two values (see \S6).

We now consider the longitude distribution of electron columns in the Galactic plane. Taking the \herschel\ points shown in figure \ref{fig:EM_GalL_HerschelPlanck} leads to the inferred DM profile shown as red points in the left panel of figure \ref{fig:DM_GalL_Herschel}. Also shown in the figure are the model thin disk DMs for both TC93 and YMW16. It is clear that (i) the N$^+$ detections are mostly confined in longitude to the same range where the model thin disk component is significant for both TC93 and YMW16, and (ii) although there are large fluctuations from point-to-point, the DM values inferred from \herschel\ FIR line intensities are in many places a substantial fraction of the DM ascribed to the thin disk.

To emphasise the second point we have also computed lower bounds on the DMs implied by the \herschel\ data, as shown in the right panel of figure \ref{fig:DM_GalL_Herschel}. The bounds are evaluated from the N$^+$ column-densities reported by G15, assuming a nitrogen abundance of $8.5\times10^{-5}$ (as per the cosmic abundance pattern specified in \chianti), and they correspond to the case where all of the nitrogen is N$^+$. As the emissivity per N$^+$ ion varies by only $\pm10\%$ over the interval $T=20{,}000\pm15{,}000\,{\rm K}$ (determined with \chianti, for both N$^+$ lines), the column-densities reported by G15 are insensitive to the assumed temperature. Although our DM lower-limits are necessarily smaller than the estimates shown in the left panel of figure \ref{fig:DM_GalL_Herschel}, the difference is modest.

The envelope of the red points in figure \ref{fig:DM_GalL_Herschel}, in either panel, is similar to the shape of the blue curves. And its overall amplitude (left-hand panel) is intermediate between the TC93 and YMW16 model thin disk components. We note the large difference in normalisation -- a factor of approximately 3.6 -- between the thin disk components in YMW16 and TC93.

Considering these points we conclude that the thin-disk component in both free-electron models is well-matched to the distribution of D-WIM, as seen with \herschel\ and \planck. The normalisation of the YMW16 thin disk is higher than can be explained with D-WIM alone, but in the case of TC93 it appears that the thin disk component could be made mainly of D-WIM.

\subsection{Spiral arms}
At high latitudes the spiral arm DM contributions are very different in YMW16 and TC93: they are $1.9\,{\rm pc\,cm^{-3}}$, and $\sim10^{-10}\,{\rm pc\,cm^{-3}}$, respectively. Thus the implied EM values, if the spiral arm free electrons are made of D-WIM, would be $\sim60\,{\rm pc\,cm^{-6}}$, which is sixty times too large, or $\sim3\times10^{-9}\,{\rm pc\,cm^{-6}}$, which is negligibly small relative to the observed value. 

\begin{figure}
\includegraphics[width=85mm]{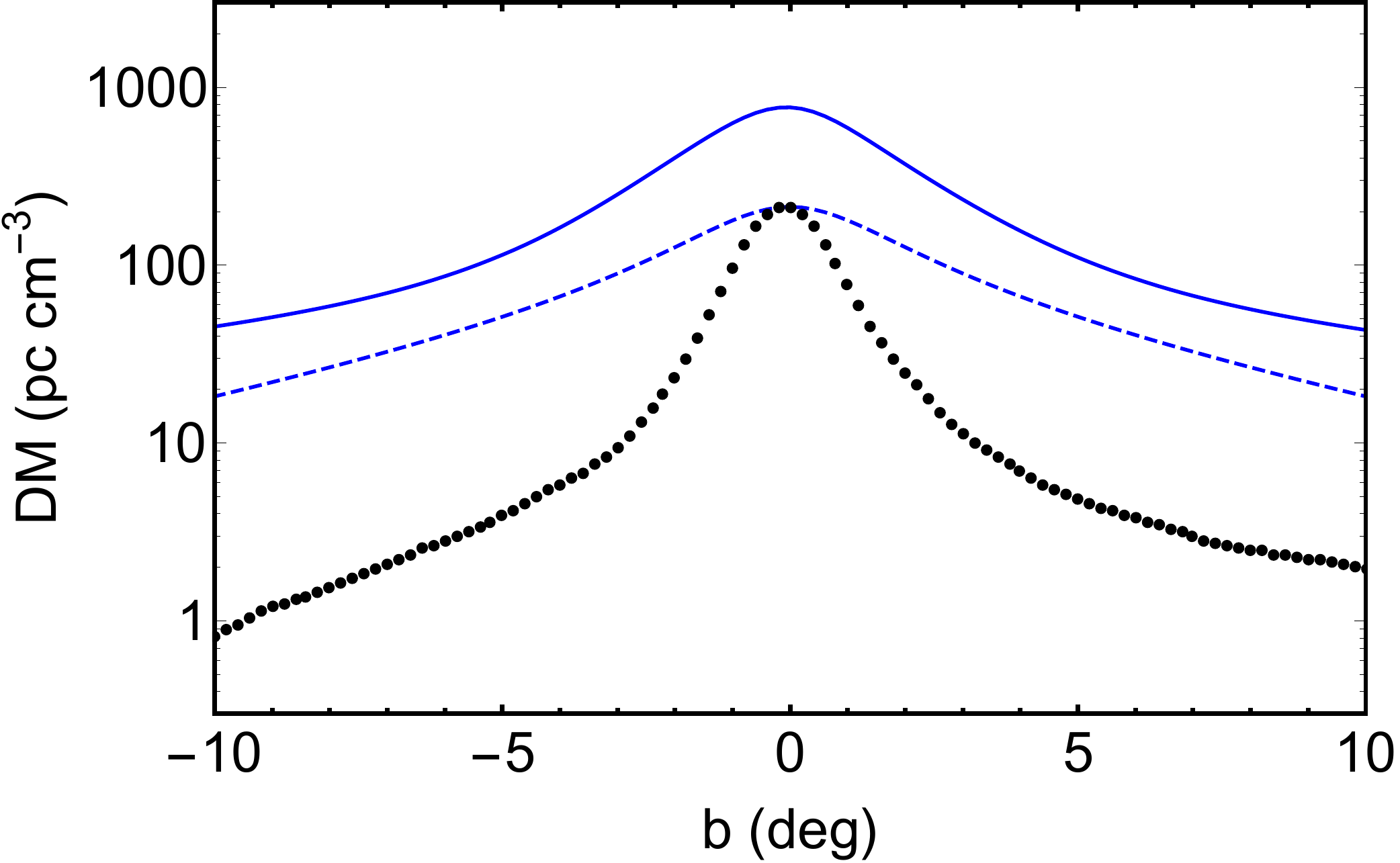}
\vskip-0.1truecm
\caption{DM as a function of Galactic latitude. Black points are the \planck\ bremsstrahlung component as per figure \ref{fig:DM_GalB_Planck_thindisk}. Also shown are the spiral arm components of the TC93 (dashed blue curve) and YMW16 (solid blue curve) free electron models.}
\label{fig:DM_GalB_Planck_spiralarms}
\end{figure}

In their vertical distribution we expect the spiral arm components of both models to be too large to match the \planck\ data, partly because the scale-heights are greater than those of the thin disk, and partly because of the proximity of some of the spiral structure to us. That expectation is borne out in practice, as can be seen from the latitude distribution shown in figure \ref{fig:DM_GalB_Planck_spiralarms}: both models show slower declines with latitude than is seen in emission, with the result that they are both at least ten times higher than the data at $|b|\simeq10^\circ$.  

\begin{figure}
\includegraphics[width=85mm]{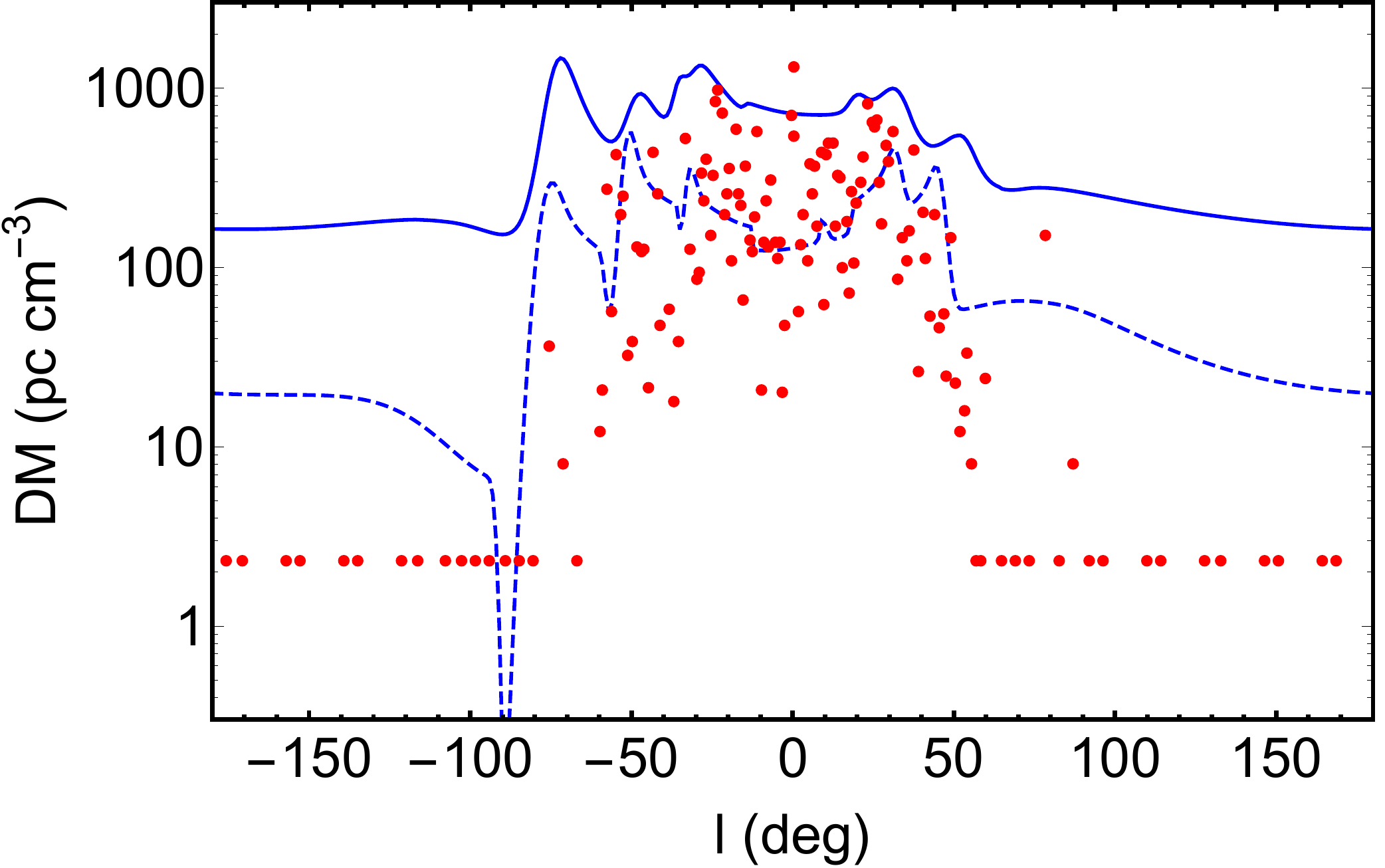}
\vskip-0.1truecm
\caption{DM as a function of Galactic longitude. Red points are the same as shown in the left panel of figure \ref{fig:DM_GalL_Herschel}. Also shown are the spiral arm components of the TC93 (dashed blue curve) and YMW16 (solid blue curve) free electron models.}
\label{fig:DM_GalL_Planck_spiralarms}
\end{figure}

The longitude distribution of the model spiral arm components is shown in figure \ref{fig:DM_GalL_Planck_spiralarms}, along with the \herschel\ data. As with the thin disk model components, the spiral arm contribution exhibits a concentration toward the inner Galaxy. However, the concentration in the free-electron models is not as strong as in the \herschel\ data (or the \planck\ data). In the case of TC93, the spiral arm DMs in the outer Galactic disk exhibit an order of magnitude more DM than could be supplied by the D-WIM, and for YMW16 the discrepancy is two orders of magnitude. We conclude that both longitude and latitude distributions of the spiral arm components of TC93 and YMW16 are inconsistent with them being made of D-WIM.

\section{D-WIM in the Solar neighbourhood} 
The fact that the EM of the WIM in the Galactic plane is dominated by the D-WIM naturally raises the question ``how much does dense plasma contribute to the emissions observed at high latitudes?'' In other words, how important is the D-WIM locally? In the previous section we concluded that the D-WIM observed by G15 should be identified with the thin disk component in the free-electron models of YMW16 and TC93. Those models yield predictions for the whole sky, so they provide an answer of sorts. As noted in \S5.3, both YMW16 and TC93 imply EM contributions at high latitude that are interestingly large --- comparable to the values (${\rm EM}\sim1\,{\rm pc\,cm^{-6}}$) usually attributed to the WIM, in fact. In this section we consider three possible observational manifestations of D-WIM at high latitude.\footnote{We do not include, here, the contribution of D-WIM to pulse dispersion at high latitudes, which was considered in \S5.1 and found to be negligible.}

\subsection{Microwaves and H-alpha}
Several detailed studies of the relationship between \ha\ and the free-free microwave continuum have previously been undertaken \citep[e.g.][]{davies2006,ade2016}. Here we again make use of the \planck\ free-free map, whose high-latitude structure has been shown  to be proportional to the \ha\ intensity, with a constant of proportionality $\eta=8\pm1\,{\rm \mu K\,R^{-1}}$ at a radio frequency of $22.8\,{\rm GHz}$ \citep{ade2016}.  

\begin{figure}
\includegraphics[width=85mm]{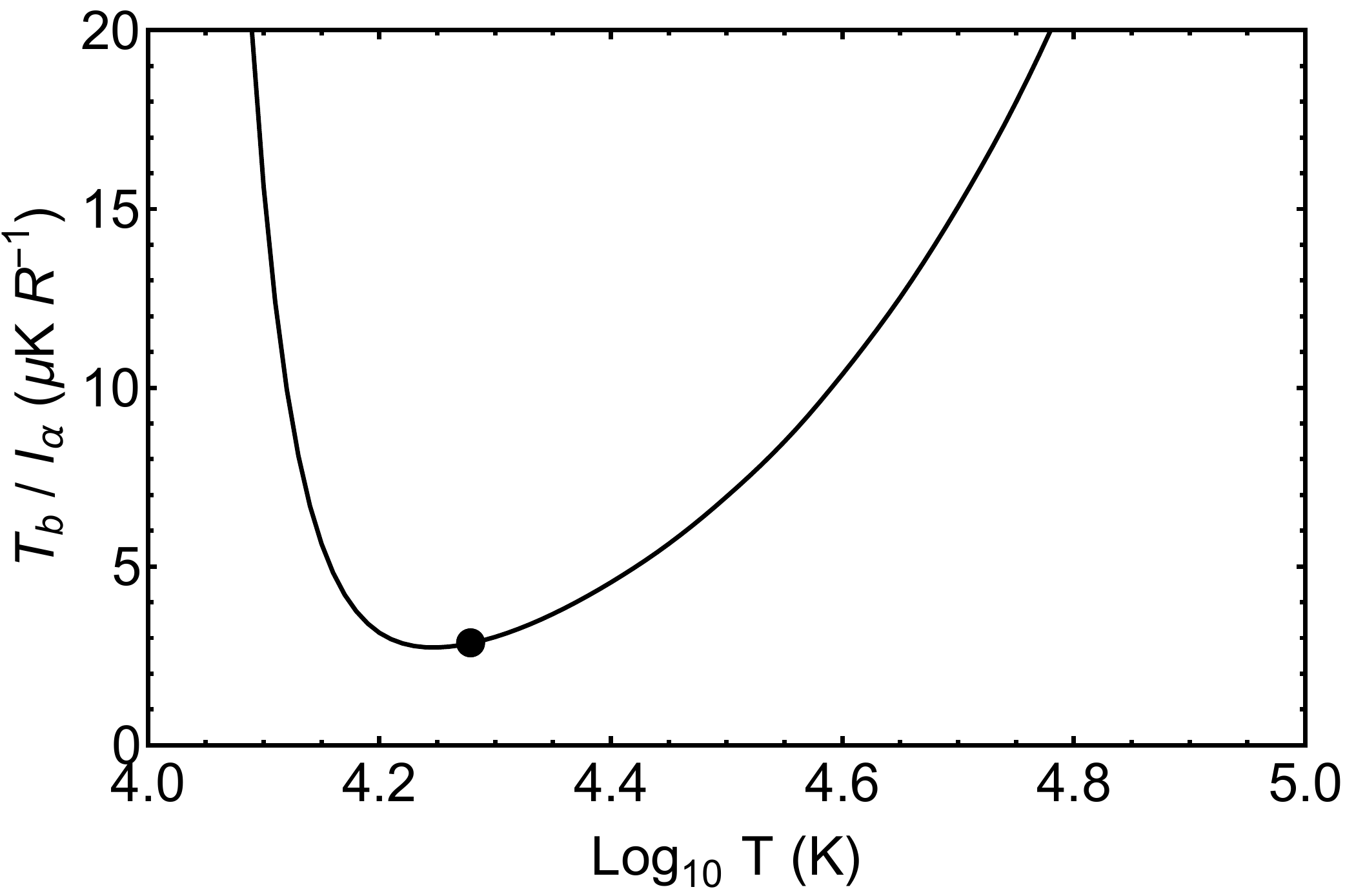}
\vskip-0.1truecm
\caption{The microwave brightness at $22.8\,{\rm GHz}$, per unit \ha\ intensity, as a function of temperature for our plasma model (\S2.2), with $n_e=30\,{\rm cm^{-3}}$. The black dot corresponds to our preferred D-WIM temperature for the inner Galactic plane of $19{,}000\;{\rm K}$ (\S\S3.2,4.2).}
\label{fig:hamicro_22.8GHz}
\end{figure}

For our plasma model, the microwave brightness per unit \ha\ intensity is as shown in figure \ref{fig:hamicro_22.8GHz} where we see that $\eta$ decreases with temperature up to about $17{,}000\,{\rm K}$, increasing thereafter. This behaviour is quite different to that of photoionised plasma \citep[e.g.][]{dong2011}, for which $\eta$ increases monotonically with temperature. The difference is readily understood: at low temperature, a collisional ionisation equilibrium favours the production of ions from elements that have lower ionisation potentials than hydrogen, yielding weaker \ha\ emission as the temperature drops. By contrast, if there are Lyman continuum photons and hydrogen atoms, as in a photoionised plasma, then ionisation of hydrogen will take place even at low temperatures, and that leads to more \ha\ as recombination is faster at lower temperatures. (These dependencies are stronger than the temperature variation of the free-free microwave continuum, which scales approximately as $1/\sqrt{T}$.)

\begin{figure}
\includegraphics[width=80mm]{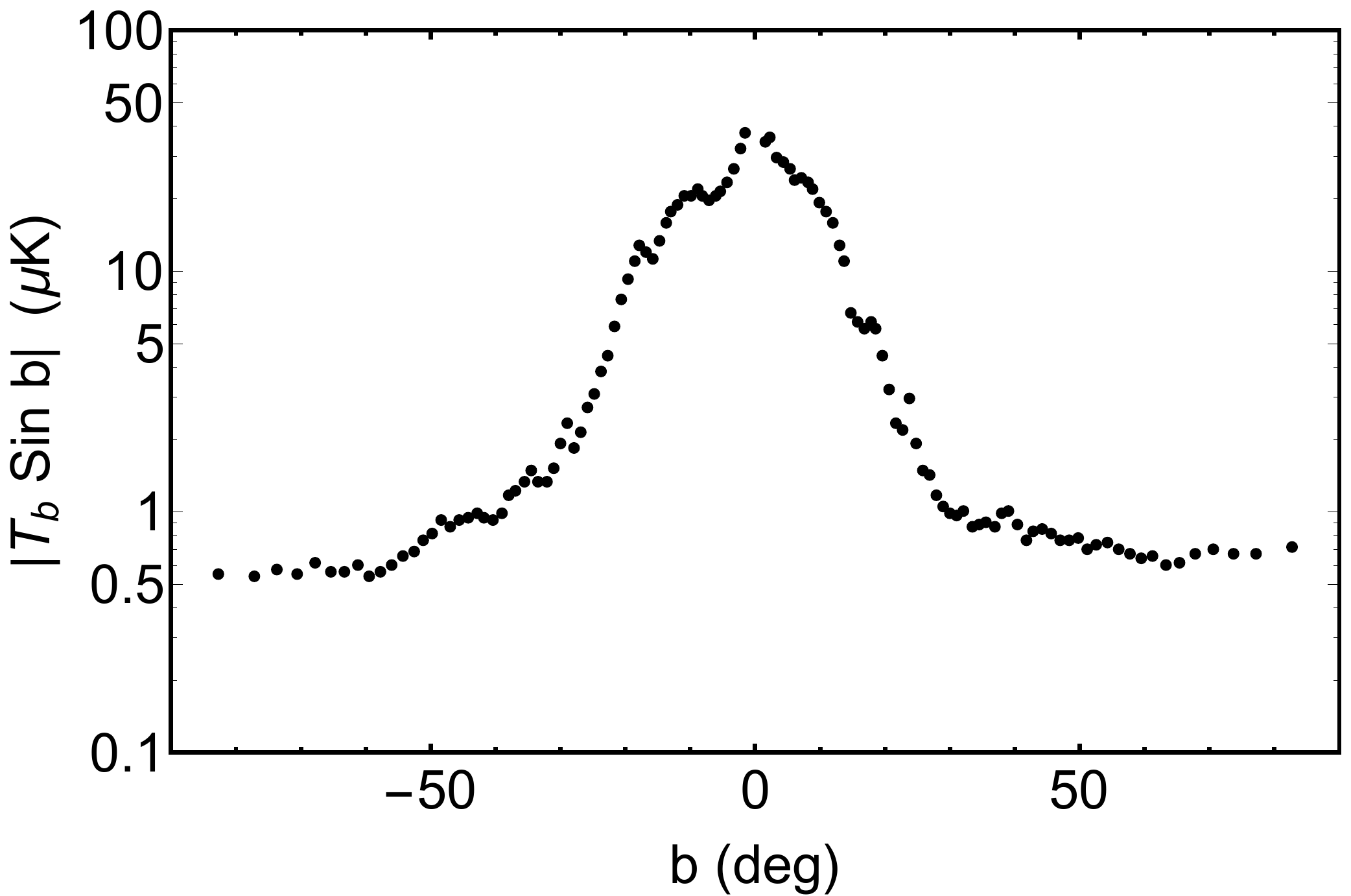}
\includegraphics[width=80mm]{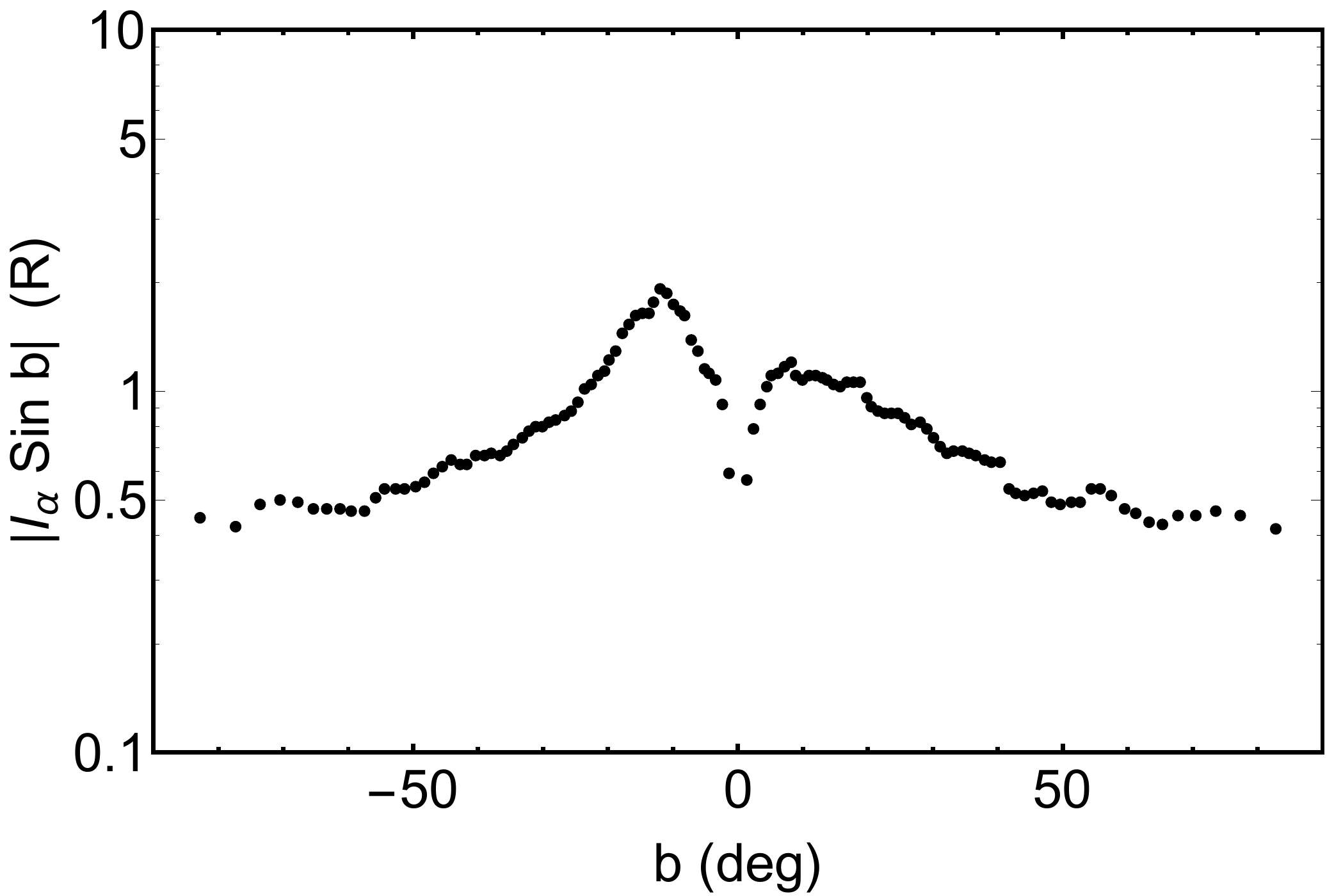}
\includegraphics[width=80mm]{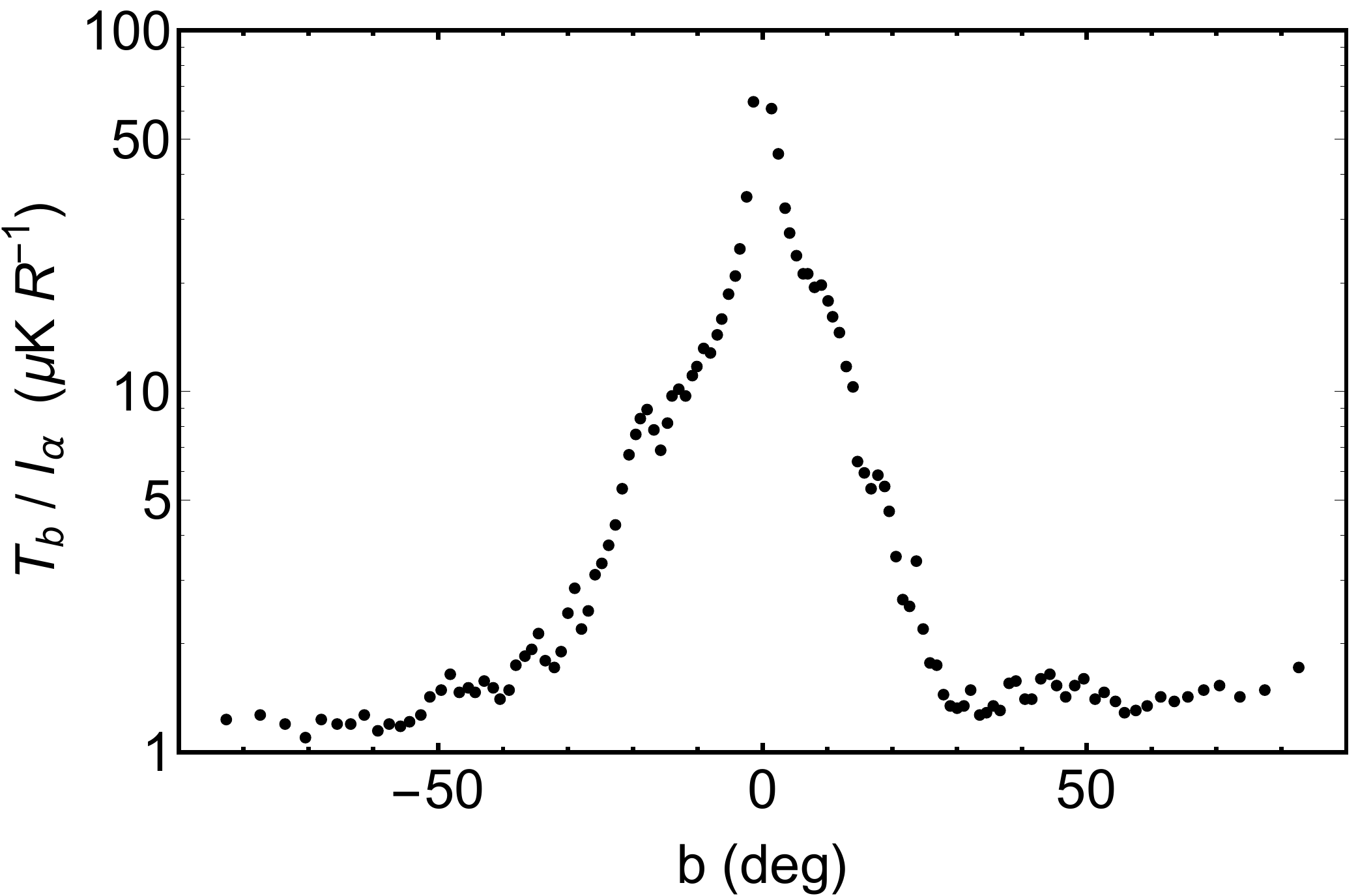}
\vskip-0.1truecm
\caption{Median statistics of the \planck\ free-free brightness temperature map, evaluated at $22.8\,{\rm GHz}$ (top panel); the \ha\ intensity (middle panel); and the ratio of the two (bottom panel). }
\label{fig:median_freefree_22.8GHz}
\end{figure}

To match the reported value of $\eta=8\pm1\,{\rm \mu K\,R^{-1}}$ with our plasma model would require $T=13{,}500\pm200\,{\rm K}$, or  $T=34{,}000\pm3{,}000\,{\rm K}$. Of these two, the latter can be excluded by considering other optical emission lines, as described in the next section. Although $T=13{,}500\,{\rm K}$ is hot for a photoionised plasma, it is substantially cooler than the value of $19{,}000\,{\rm K}$ we deduced earlier for the D-WIM in the inner Galaxy (\S\S3.2,4.2). However, it is important to note that the proportionality between free-free and \ha\ emission that was demonstrated by \citet{ade2016} relies on high signal levels --- see particularly their figure 4, which extends up to $200\,{\rm R}$ in \ha\ intensity. Bearing in mind that only $\sim1\,{\rm R}$ is attributed to the WIM at high latitudes, the constant of proportionality established by \citet{ade2016} does not necessarily apply to the WIM.

To underline the point just made, we refer to figures 5 and 6 of \citet{haffner1999} \citep[see also][]{madsen2006} which demonstrate clear trends in the optical line ratios with \ha\ intensity. The same authors attributed those trends to temperature variations, with lower \ha\ intensity corresponding to hotter gas. We therefore consider the possibility that $\eta$ might also change with signal level, and we attempt to determine its value at low \ha\ intensities.

\subsubsection{Weak free-free continuum at low H$\alpha$ intensities}
When the signal levels are low, it is helpful to average the data. Taking mean values is not appropriate, as means are often dominated by the small fraction of data exhibiting strong signals --- certainly that is the case for the \planck\ free-free map at high latitudes, where point sources (flat-spectrum radio quasars) dominate the mean. Instead we take the median values of both \ha\ \citep[from][]{finkbeiner2003} and microwave free-free emission \citep[from][]{adam2016} over annular regions defined by lines of constant Galactic latitude, and extending over all longitudes. The results are shown in figure \ref{fig:median_freefree_22.8GHz}, along with the ratio of free-free to \ha\ intensity.

The qualitative behaviour seen in the top two panels of figure \ref{fig:median_freefree_22.8GHz} is unsurprising. At low latitudes the microwave signal rises strongly as one approaches the plane, despite the moderating influence of the $|\sin b\,|$ factor. The \ha\ signal, on the other hand, first increases and then declines as dust extinction limits the distance to which we can see in the optical. At high latitudes microwave and \ha\ signals both vary roughly in proportion to $1/|\sin b\,|$, as expected for a distribution that is locally approximately plane-parallel. What is remarkable, however, is the very low ratio of the microwave free-free continuum to \ha\ intensity obtained at high latitude, as shown in the bottom panel of the figure. Based on figure \ref{fig:hamicro_22.8GHz}, and the temperature of $19{,}000\,{\rm K}$ we deduced earlier (\S\S3.2,4.2) we expect a ratio $\eta\simeq3\;{\rm \mu K\,R^{-1}}$, whereas at high latitudes we observe a value approximately half that. Moreover, at our preferred temperature the theoretical curve is close to its minimum, so we cannot align model and data by choosing a different temperature. 

Part of the discrepancy is presumably due to scattered light contributing to the measured \ha\ at high latitude. However, if we adopt a scattered fraction of $\simeq20$\% \citep[e.g.][]{brandt2012}, a substantial discrepancy remains. Eliminating the discrepancy would require a scattered \ha\ fraction of  $\simeq50$\% --- a high value, but one that has previously been suggested by other authors for different reasons \citep{witt2010}.

Alternatively, it is possible that at low signal levels the spectral decomposition of the \planck\ foregrounds is not as reliable as it appears to be at high signal levels, so that a significant fraction of the free-free intensity has been incorrectly assigned to the Synchrotron or Anomalous Microwave Emission channels.

Although there is a discrepancy between the typical (i.e. using medians) high latitude value of $\eta$ (figure 11) and our model expectation, the disagreement is modest. Simple photoionisation models, by contrast, already encounter difficulties with the value $\eta\simeq8\;{\rm \mu K\,R^{-1}}$ inferred for high intensities, and those problems become more severe as $\eta$ decreases.

\subsection{Optical emission line spectra}
Various optical emission lines have been observed in the local diffuse light, and can be used to constrain the properties of the WIM \citep[e.g.][]{reynolds2004,madsen2006,haffner2009}. Compared to the optical spectra of classical HII regions -- i.e. the photoionised nebulae surrounding O and B stars -- the optical spectrum of the WIM is weak in [OIII] and HeI, but strong in [NII] and [SII].

\begin{figure*}
\includegraphics[width=85mm]{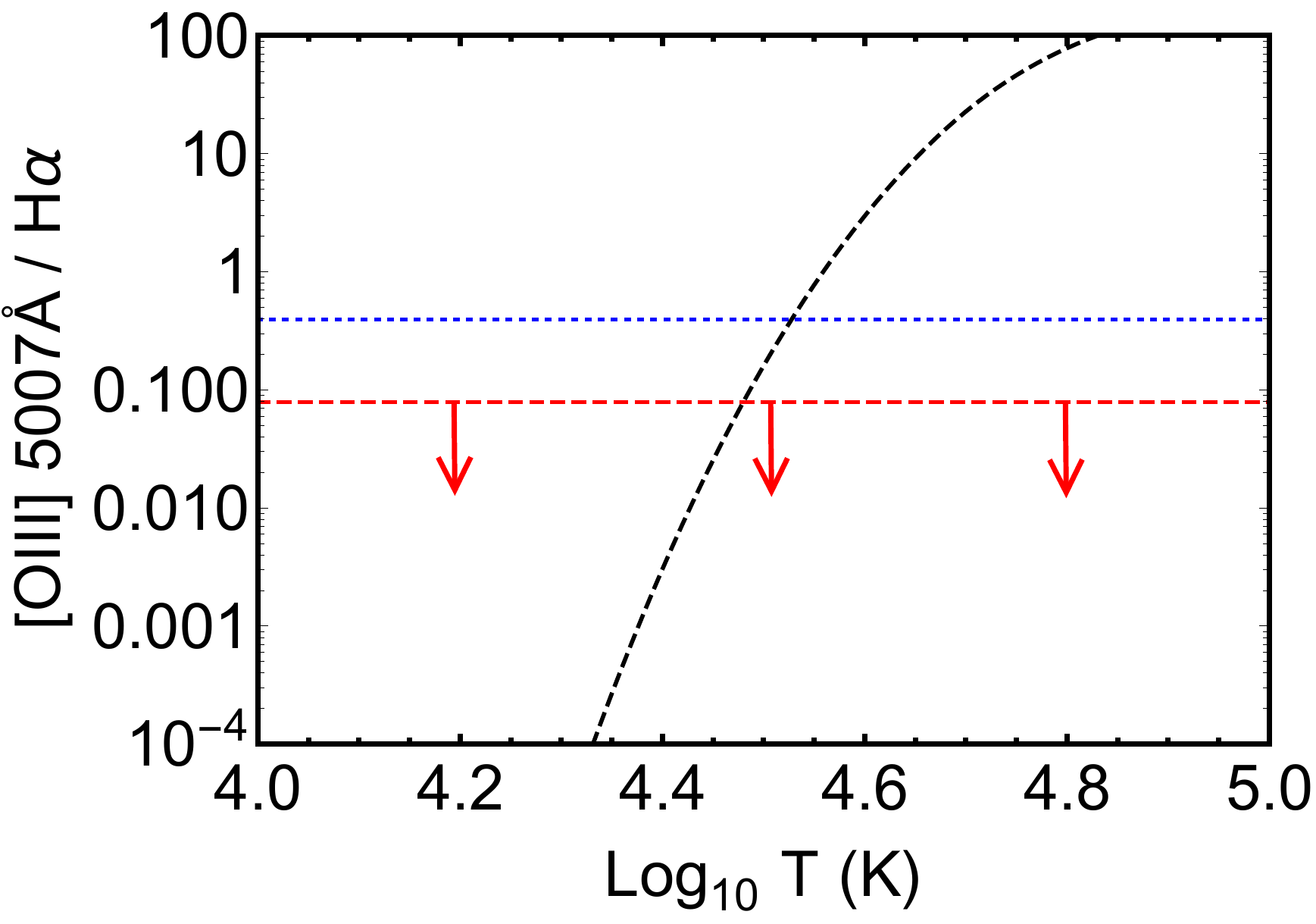}\ \ \includegraphics[width=85mm]{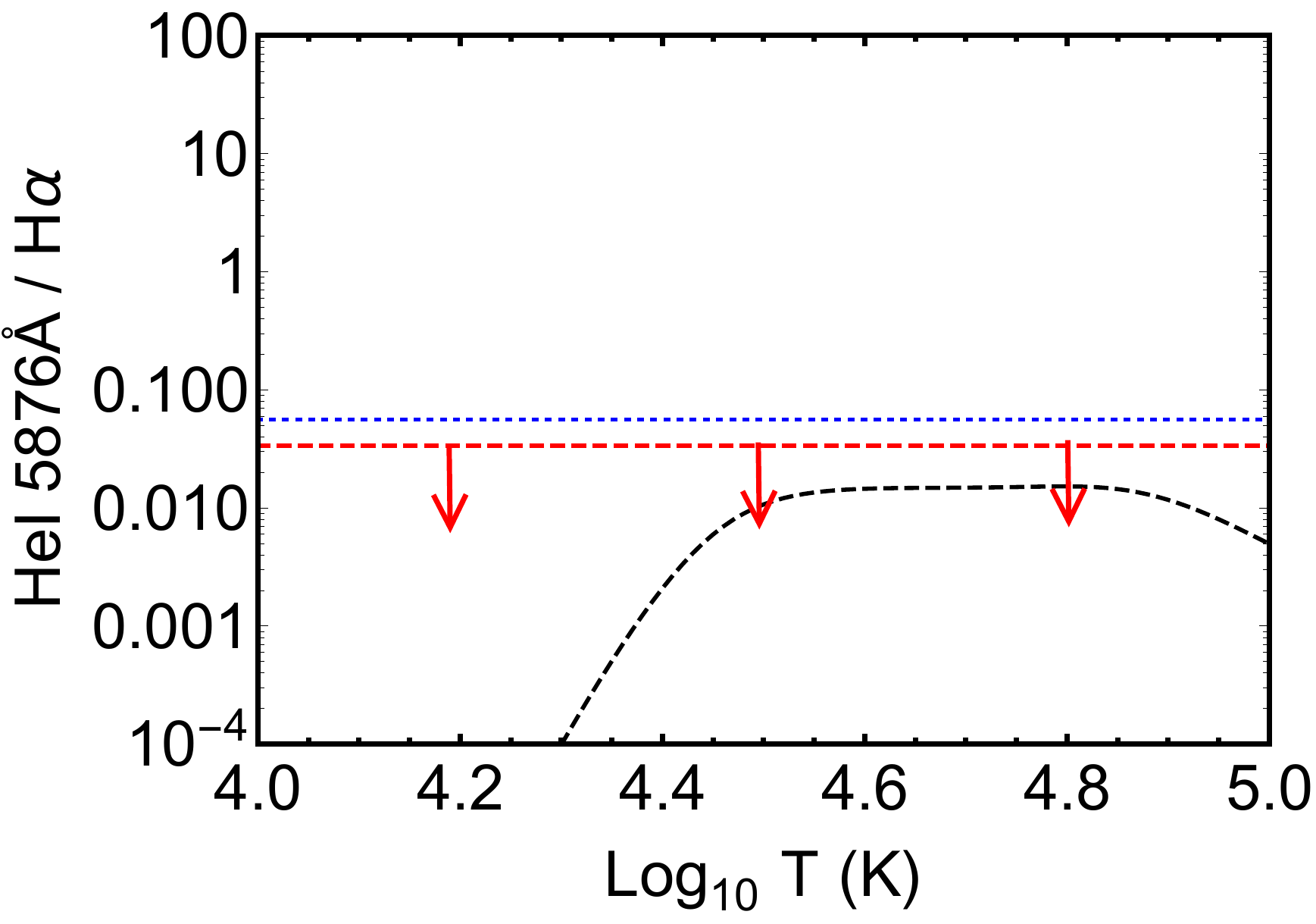}
\includegraphics[width=85mm]{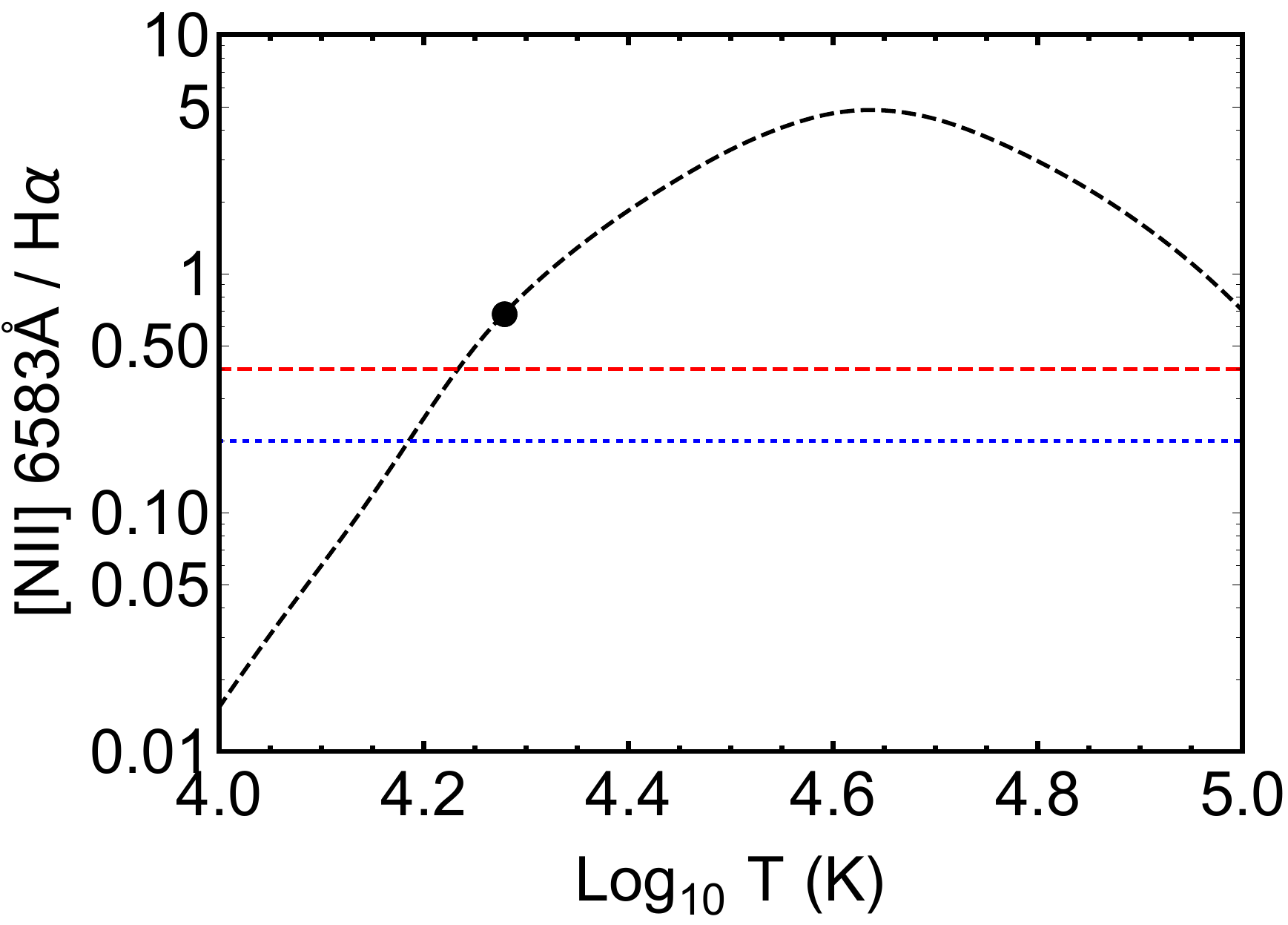}\ \ \includegraphics[width=85mm]{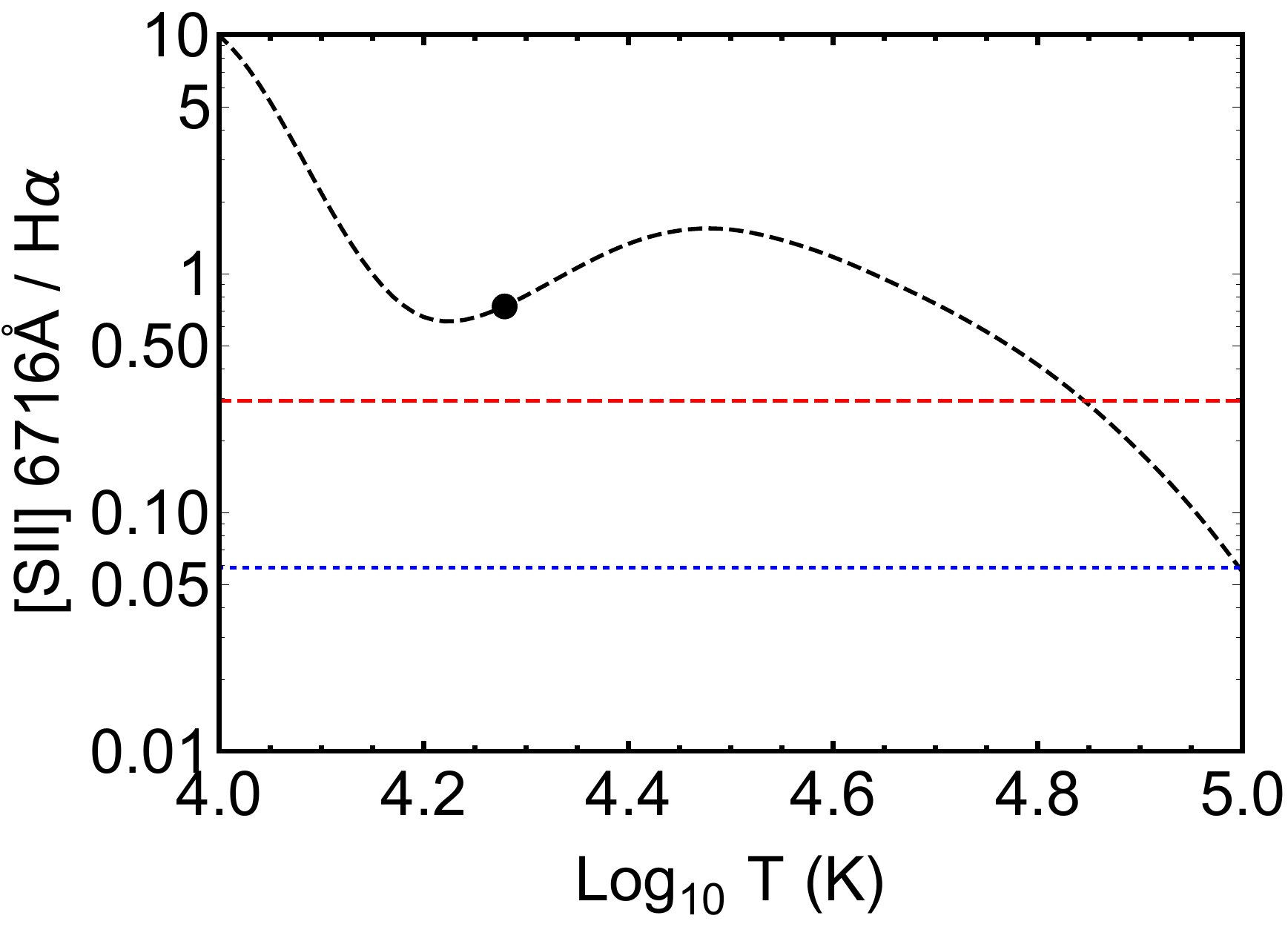}
\includegraphics[width=85mm]{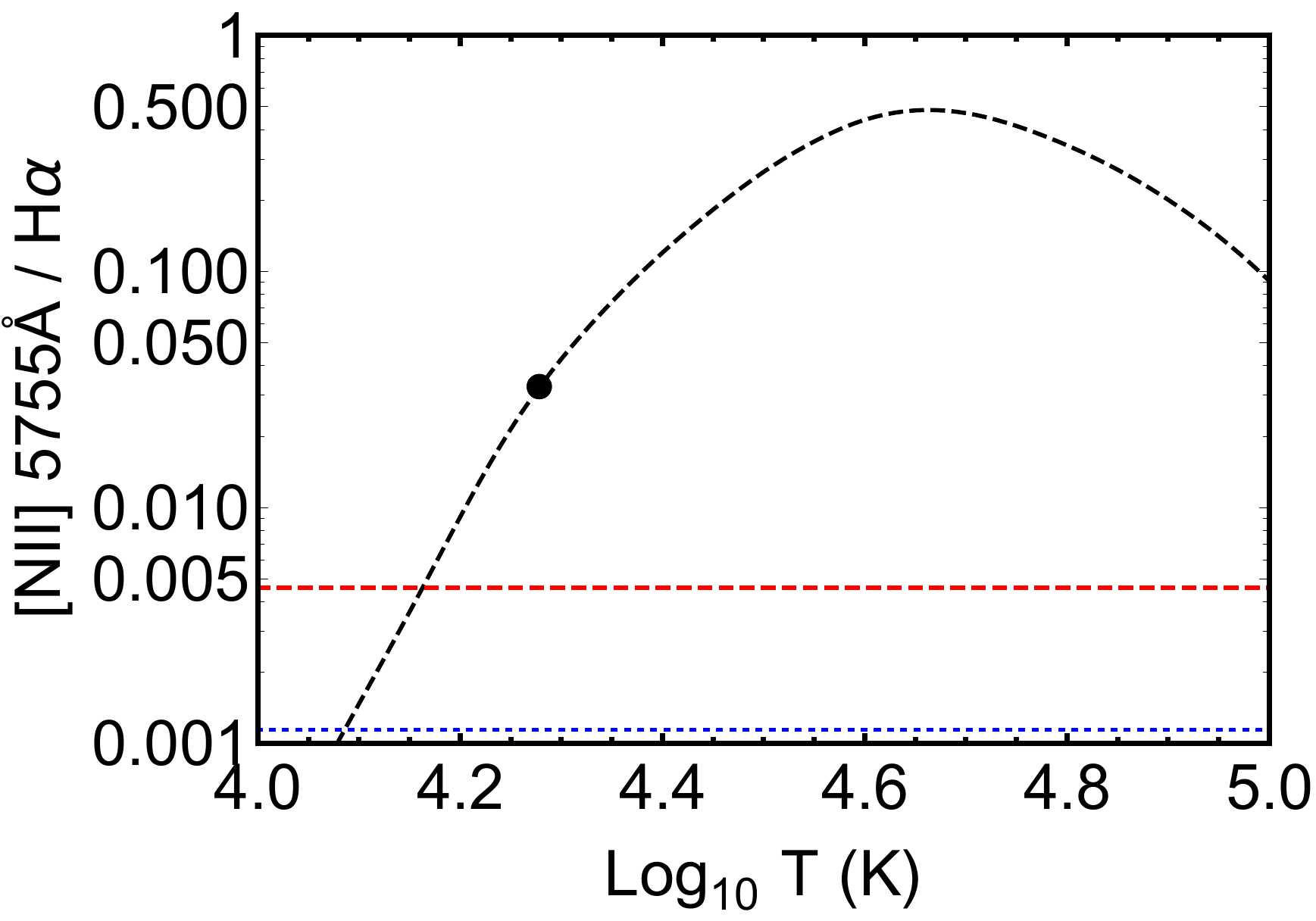}\ \ \includegraphics[width=85mm]{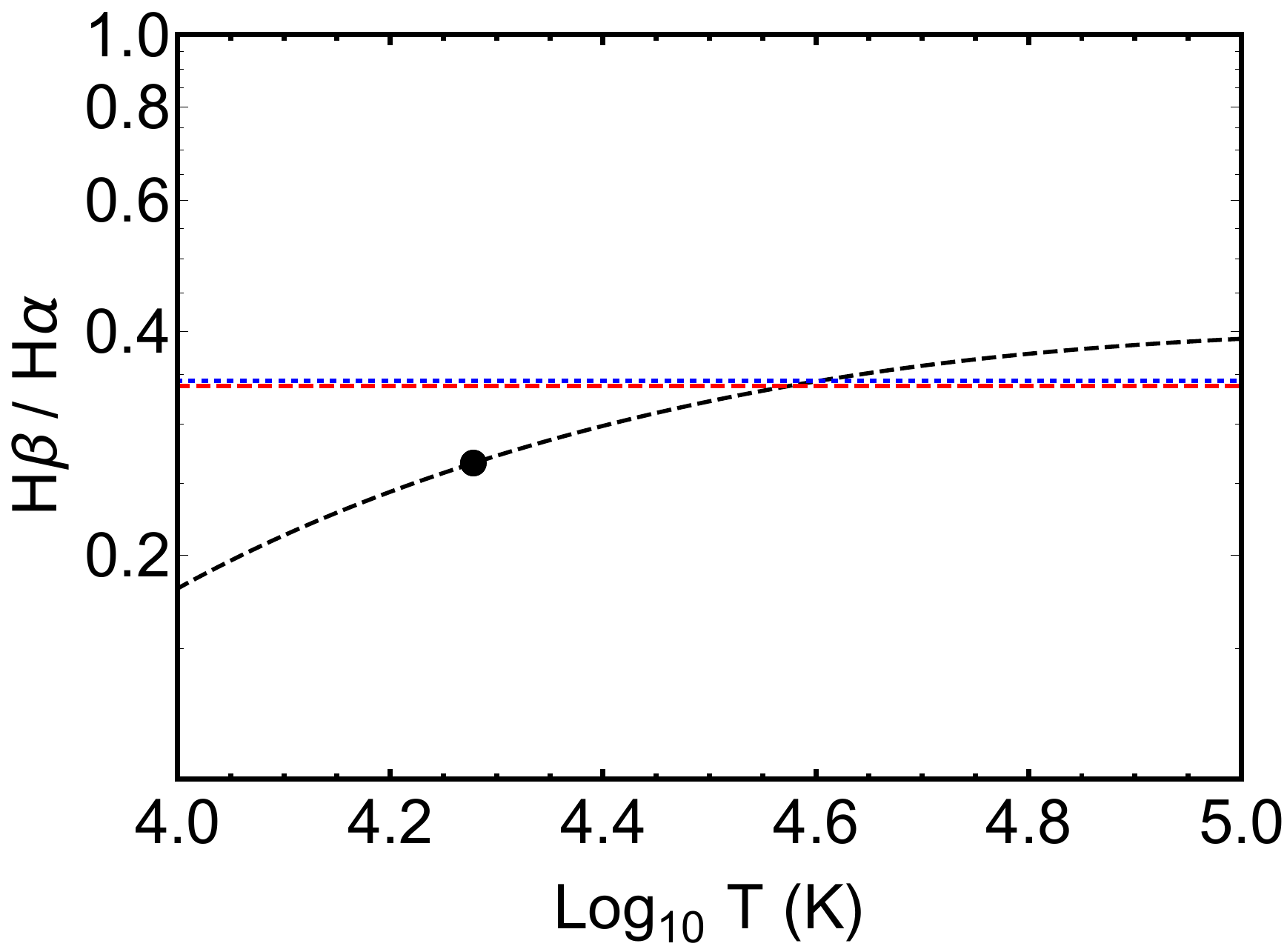}
\vskip-0.1truecm
\caption{Optical emission line strengths (in energy units), relative to \ha, for the six transitions given by \citet{reynolds2004}. The dashed black lines show the predictions of our plasma model, for $n_e=n_g$, as a function of temperature. The black dot corresponds to our preferred D-WIM temperature for the inner Galactic plane of $19{,}000\;{\rm K}$ (\S\S3.2,4.2). Dashed red and blue lines are the line ratios for the WIM and for bright HII regions, respectively, as given by \citet{reynolds2004}.}
\label{fig:optical}
\end{figure*}

Once the temperature and density are specified, our plasma model makes predictions for emission line strengths. We take $n_e=n_g$ for the D-WIM, and calculate line ratios as a function of temperature. We have considered the six lines whose typical strengths were given by \citet{reynolds2004}, with the results shown in figure \ref{fig:optical} (dashed black lines) for [NII] (6583\AA), [SII] (6716\AA), [OIII] (5007\AA), HeI (5876\AA), [NII] (5755\AA), and H$\beta$. For comparison we also plot typical values for classical HII regions (dashed blue lines) and fiducial values for the WIM (dashed red lines), as given by \citet{reynolds2004}.

The notional WIM spectrum given by \citet{reynolds2004} is not the spectrum of the WIM alone. The WIM is certainly a major contributor, but there is also expected to be light that is scattered off dust. The dust-scattered light was estimated by \citet{brandt2012} to be $19\pm4$\% of the total \ha\ intensity, on average at high latitude, but larger values are possible \citep{witt2010}. Scattered light presumably includes a substantial contribution from classical HII regions in the Galactic plane, as they are very bright, and as a result we expect that the notional WIM line ratios given by \citet{reynolds2004} will be intermediate between those of an HII region and those of the true WIM spectrum.

For the [OIII] and HeI lines (top two panels of figure \ref{fig:optical}), which are both strong in HII regions, and which are blueward of \ha\ so the percentage of scattered light should be greater, the level of contamination is presumably several times larger than the $\sim20$\% estimated for \ha. We therefore expect that the true [OIII] and HeI line strengths of the WIM are much weaker than the notional WIM values given by \citet{reynolds2004}, and we conclude that the latter should be treated as upper limits. (Nevertheless the requirement for very low levels of [OIII] emission from the WIM can still be useful: it excludes the hotter of the two possible solutions mentioned in \S6.1 for a microwave-to-\ha\ ratio of $\eta=8\,{\rm \mu K\, R^{-1}}$.)

For  [NII] (6583\AA) and  [SII] (6716\AA) (middle two panels of figure \ref{fig:optical}), the situation is reversed: these lines are stronger in the notional WIM spectrum (relative to \ha) than in classical HII regions. Furthermore, both lines are slightly to the red of \ha. We therefore expect that the ratios reported by \citet{reynolds2004} for these lines ought to be close to those of the WIM itself, but that the latter should be somewhat larger than the notional value in each case.   

Some further insight into the intrinsic line strengths of the WIM is provided by the results of \citet[][figures 5 \& 6]{haffner1999}, who showed that at high \ha\ intensities the [NII] (6583\AA) and  [SII] (6716\AA) line strengths, relative to \ha, tend towards the low values appropriate to an HII region, but become much larger in regions of the sky where the \ha\ emission is faint. This behaviour is consistent with the observed lines being due to a mixture of the WIM and classical HII regions, with HII regions contributing very strongly in some regions of the sky --- either because of scattering off dust, or because in some directions we are seeing HII regions directly. With that point in mind, figures 5 and 6 of \citet{haffner1999} suggest that the line-ratios intrinsic to the WIM might actually be best represented by the data at the lowest \ha\ intensities. In the case of the local plasma, that corresponds to roughly 1.5 and 1.2 times the \ha\ intensity, for [NII] (6583\AA) and [SII] (6716\AA), respectively.  (The corresponding numbers for the Perseus spiral arm are somewhat larger.)

Our model prediction, with $T=19{,}000\,{\rm K}$, is shown as a black dot in the relevant panels of figure \ref{fig:optical}. For [NII] (6583\AA) and [SII] (6716\AA) the prediction is intermediate between the values measured at low \ha\ intensity and the notional WIM spectrum of \citet{reynolds2004}.

The bottom panels of figure \ref{fig:optical} show the final two line ratios, of the six given by \citet{reynolds2004}. In the case of H$\beta$ the first point to notice is that there is very little variation in its strength, relative to H$\alpha$, amounting to only a factor of two over the whole temperature range shown, whereas all the other lines in figure \ref{fig:optical} vary by orders of magnitude over the same temperature range. Consequently, even though it is a relatively strong line, H$\beta$ is not a very powerful diagnostic of the plasma conditions. This problem is compounded by the fact that H$\beta$ is more strongly affected than \ha\ by dust scattering and by extinction, so a good description of these effects is needed in order to interpret the measured line ratio. Because of these difficulties it is unclear whether the H$\beta/$H$\alpha$ ratio offers a useful test of our model.

The situation is very different for [NII] (5755\AA), which varies in strength by three orders of magnitude between $10{,}000\;{\rm K}$ and $50{,}000\;{\rm K}$. Moreover, because we have another line from the same species, arising from a different upper level, the ratio [NII] (5755\AA)/[NII] (6583\AA) provides a direct gauge of the plasma temperature, essentially independent of the physical model under consideration. This line ratio has been used previously to infer the temperature of the WIM with values ranging up to approximately $12{,}000\,{\rm K}$ \citep{haffner2009} --- hotter than expected for a photoionised plasma, but cooler than the $19{,}000\,{\rm K}$ we deduced in \S\S3.2, 4.2. On the face of it, these observations seem to challenge our model; but the challenge is not a direct confrontation, for the following reason. The [NII] (5755\AA) is very weak in comparison with H$\alpha$, so it is only possible to measure it in regions of the sky where the H$\alpha$ is strong and, as we have already noted, there are clear trends in the line ratios that point to higher temperatures at lower signal levels \citep{haffner1999}. 

\subsection{Extreme radio-wave scattering}
As stated in the Introduction, there is a substantial similarity between the plasma properties determined by G15 and those inferred from extreme radio-wave scattering (ERS). In particular we note that (i) $n_e\sim10$-$30\,{\rm cm^{-3}}$ has been inferred for the IDV of quasars and pulsar parabolic arcs \citep{rickett2011,tuntsov2013}, and (ii) ERS phenomena are seen all over the sky, implying that the plasma clouds responsible for the scattering are commonplace in the vicinity of the Sun. Given that ERS requires widespread, high-density plasma, it is presumably a manifestation of D-WIM in the solar neighbourhood. A key question is then whether the incidence of ERS is consistent with the observed line intensities at high latitude, under the assumption of a common origin.

To make that connection we need to know the size of the individual D-WIM structures. As noted in the introduction, transience of the ERS phenomena has led to inferred sizes $\sim10^{1\pm1}\;{\rm AU}$ for the ERS clouds. We take the upper end of that range, because the lower end tends to be associated with the more extreme refractive events \citep[e.g.][]{bannister2016}, which could be simply substructure within a broader envelope of D-WIM.

The intensity of emission from the D-WIM and the geometric optical depth of the dense plasma clouds are proportional to each other (\S2.1). At $|b|>60^\circ$ the median value of $I_\alpha |\sin b\,|$ (middle panel of figure 11) is $0.46\,{\rm R}$, and for our plasma model with $n_e=30\,{\rm cm^{-3}}$ and $T=19{,}000\,{\rm K}$ that corresponds to ${\rm EM}\simeq0.36\,{\rm pc\,cm^{-6}}$. Thus if the D-WIM at high latitude is made up of clouds of size $R\sim100\,{\rm AU}$, and contributes all of the observed optical emission, then we expect an optical depth $\tau\sim0.6$. That is more than enough to explain the rate at which ERS is observed in compact radio quasars. For the most extreme IDV, the observed optical depth is only $\sim10^{-3}$ \citep{lovell2008}. That is so small that it might at first sight seem inconsistent with $\tau\sim0.6$. However, the angular size of radio quasars introduces significant smoothing into the observed light-curves, resulting in strong selection biases: fast, large-amplitude variability can only arise from very local plasma. Distances $\sim10\,{\rm pc}$ have been deduced for the two best-studied examples of extreme IDV \citep{dennettthorpe2003,bignall2003}, whereas the scale-height of the material responsible for the optical emission lines is $\sim1\,{\rm kpc}$ \citep{haffner1999}. Consequently the IDV phenomenon is insensitive to the bulk of the emitting plasma, even if that plasma is D-WIM. 

Radio pulsars have much higher brightness temperatures than radio quasars, and are therefore less susceptible to the selection bias just discussed. The pulsar counterpart to IDV in quasars is the parabolic arc phenomenon \citep{stinebring2001,cordes2006,rickett2011,tuntsov2013}. Parabolic arcs appear quite commonly in observations of nearby pulsars: \citet{putneystinebring2005} report a total of 20 distinct arcs in a sample of six pulsars, with a combined path of $4.4\,{\rm kpc}$, implying an optical depth per unit pathlength of $\sim4\,{\rm kpc^{-1}}$ in the immediate vicinity of the Sun. An effective pathlength of $\sim1\,{\rm kpc}$, appropriate to the WIM seen in emission at high latitude, then yields $\tau\sim4$ and ${\rm EM}\sim2\,{\rm pc\,cm^{-3}}$. These numbers are likely overestimates, as \citet{putneystinebring2005} reported on only half of their sample of pulsars, and the objects they selected were chosen because each showed multiple parabolic arcs. We conclude that the incidence of ERS in the solar neighbourhood is broadly consistent with it arising in the same gas that is responsible for the observed emission lines.

\section{Discussion}

\subsection{Validity of our adopted plasma model}
\begin{figure}
\includegraphics[width=85mm]{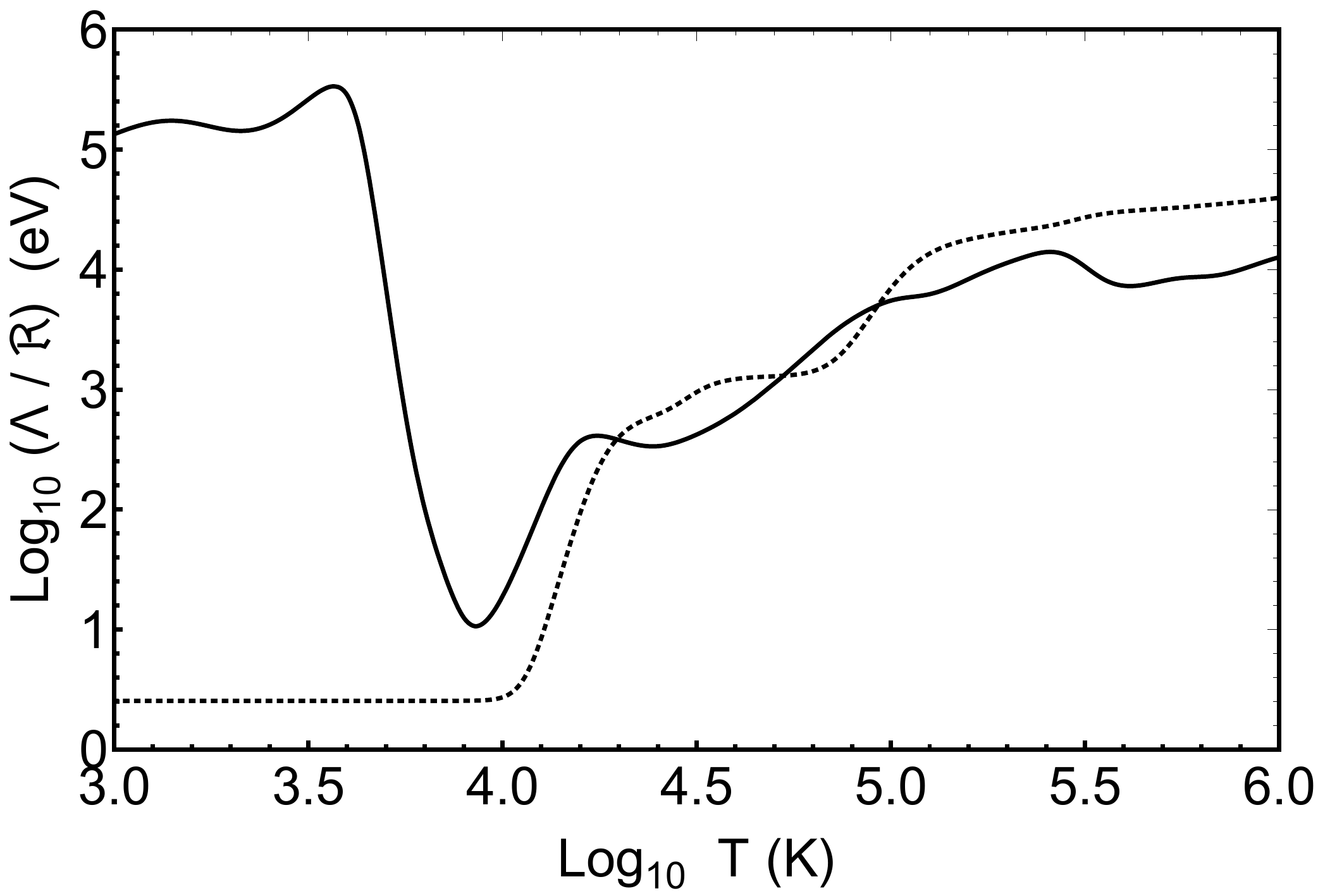}
\vskip-0.1truecm
\caption{The heat per ionisation for the plasma model used in this paper, as a function of temperature, with $n_e=30\;{\rm cm^{-3}}$ (solid line). This quantity is evaluated within \chianti\ as the total radiative cooling rate $\Lambda\;({\rm eV\,cm^{-3}\,s^{-1}})$ divided by the total recombination rate, ${\cal R}\;({\rm cm^{-3}\,s^{-1}})$. Also shown (dotted line) is a simple estimate of the heat per ionisation appropriate to cosmic-ray energisation (see \S7.1).}
\label{fig:energyperionisation}
\end{figure}

In this paper we have used a collisional ionisation equilibrium model. This corresponds to the limiting case where the energising process delivers heat to the gas but does not cause any direct ionisations, so that the heat delivered per ionisation is infinite. In steady state, the heating rate of the energising process(es) must balance the radiative cooling rate, and the ionisation rate must balance the recombination rate. We can therefore gauge whether our plasma model is likely to be a good approximation by evaluating the model radiative cooling power per recombination rate. That quantity is shown in figure \ref{fig:energyperionisation}. Any energisation processes that have a comparable, or larger heat per ionisation should be reasonably well approximated by our model.

It is clear from figure \ref{fig:energyperionisation} that our model does not offer a good approximation for UV-photoionised plasmas, for the following reason. Ultraviolet radiation fields typically decline steeply with increasing photon energy, so that almost all photoionisations are due to photons just above the ionisation edge. The average heat supplied per ionisation is then $\sim1\,{\rm eV}$ --- an order of magnitude below the lowest point of the curve in figure \ref{fig:energyperionisation}.

Our model is more relevant to the case of cosmic-ray energisation, which can supply much more heat than UV photoionisation. In a completely neutral gas, cosmic rays undergo Coulomb collisions with bound electrons and thus lose energy, causing ionisation in the process. For atomic hydrogen, on average $36\,{\rm eV}$ is lost per primary ionisation, decreasing to $23\,{\rm eV}$ per ionisation after allowing for secondary ionisations, and approximately 11\% is deposited as heat \citep[e.g][]{dalgarno1999}. Thus in a completely neutral gas there is only a few ${\rm eV}$ of heat per ionisation. At the other extreme of a fully ionised plasma, cosmic-rays lose energy via Coulomb collisions with the free electrons at a rate that is roughly five times larger than they would to bound electrons, see figures 2 and 3 of \citet{walker2016}. Furthermore in the fully ionised case the energy loss of the cosmic-rays goes entirely into heat, and the ionisation rate is zero. (So the heat per ionisation is infinite.) In the more general case of a partially ionised gas we can make a simple estimate of the heating rate per unit volume by adding the power per bound electron, weighted by the number density of bound electrons, to the power per free electron, weighted by the number density of free electrons. Only the bound electrons contribute to the ionisation rate per unit volume. And the ratio of these quantities gives us the heat per ionisation. Treating {\it all\/} bound electrons as if they were incorporated in hydrogen atoms then leads to the crude estimate shown as the dotted line in figure \ref{fig:energyperionisation}. The estimate is biased low because (i) in reality many of the bound electrons would take much more than $36\,{\rm eV}$, on average, to liberate, and (ii) when free electrons are present a greater fraction of the energy that is deposited goes into heat. Nevertheless, at temperatures above $10^4\,{\rm K}$ our estimate of the cosmic-ray heating per ionisation is nowhere far below the value for our plasma model, and we conclude that our model may serve as a useful approximation for cosmic-ray energised plasmas.

The requirement for a large heat per ionisation is of course met by any process which injects energy in quanta that are smaller than the ionisation energy. One such process that is of particular interest (see \S7.3) is the UV pumping of molecular hydrogen by photons of energy less than $13.6\,{\rm eV}$: such photons will ionise neither atomic nor molecular hydrogen, but they can drive the H$_2$ into highly excited rotation/vibration states. Radiative de-excitation of these states is very slow, because the transitions are quadrupolar, whereas collision rates are high at high densities and they drive the kinetic temperature into equilibrium with the \htwo\ rotational/vibrational distributions. Furthermore, a significant fraction \citep[$\sim10$\%, e.g.][]{draine2011} of the UV excitations result in direct dissociation of the H$_2$, with $\sim10\;{\rm eV}$ of the UV photon energy going promptly into the thermal pool. Although this is a promising method of generating a plasma in collisional ionisation equilibrium, we caution that our model plasma includes no molecular hydrogen component.

\subsection{Free electron distributions}
The free electron models considered in \S5 of this paper assign the pulse-dispersing plasma to a small number of preconceived structures in the Galaxy. Prior information is needed to construct such models, because the measured DMs carry no information as to the distribution of the plasma along the line of sight to the pulsar. Having determined that the D-WIM contributes significantly to pulse dispersion in the inner Galactic plane, and identified the (axisymmetric) thin disk of TC93 and YMW16 as the corresponding structure, we are now in a position to check whether the free electron model distributions are accurate, by using the velocity information in the D-WIM spectral lines.

There are two relevant sets of lines: the FIR forbidden lines of C$^+$ and N$^+$, and the hydrogen RRL lines. In this paper we have not made use of the RRL data, for the reason noted in the Introduction: the lines are maser lines, which makes it difficult to quantitatively interpret the measured intensities. However, the line-of-sight integrated recombination rate is proportional to the plasma emission measure so, like the FIR forbidden lines and the free-free continuum, we expect the RRLs to be dominated by the D-WIM. In the inner Galactic plane the radial velocity distribution of the RRL intensity is quite narrow at any given longitude \citep{alves2015}, suggesting that the D-WIM is distributed in a predominantly non-axisymmetric form.\footnote{An alternative interpretation is possible: the distribution of the D-WIM might be axisymmetric, but with the intensity of the RRLs strongly enhanced at the tangent points by large maser gain due to the line-of-sight velocity coherence at those points.} The areal coverage of the inner Galaxy in FIR lines is much poorer than that of the RRL surveys, but in one case at least the data show that the D-WIM is associated with a known spiral arm \citep{langer2017}. It seems, therefore, that the large scale distribution of the D-WIM in the inner Galaxy probably has spiral structure.

Future models of the Galactic free-electron distribution could benefit by using information obtained from the spectral lines emitted by the D-WIM. Of particular value would be data from the proposed FIRSPEX satellite \citep{rigopoulou2016}, which aims to map the N$^+\;(205\,{\rm \mu m})$ and C$^+\;(158\,{\rm \mu m})$ lines with high sensitivity and good angular and velocity resolution at low- and mid-latitudes.\footnote{Lines of neutral carbon and CO will also be surveyed.}

\subsection{Circumstellar clouds of D-WIM}
As discussed in \S6.3, it is likely that the plasma responsible for extreme radio wave scattering is just a local manifestation of the D-WIM identified by G15. As such, our understanding of the D-WIM benefits from studies of extreme scattering. Particularly useful for building a picture of the D-WIM is the transient nature of the various extreme scattering phenomena, which implies (\S6.3) that the scattering plasma -- and therefore presumably the D-WIM as a whole -- occurs in the form of large numbers of tiny ($\sim100\,{\rm AU}$) plasma clouds. This insight would have been difficult to obtain with observations of the D-WIM in emission alone.

We can go further. Last year evidence emerged that the plasma clouds responsible for extreme scattering are not randomly distributed, interstellar structures, as had previously been assumed; rather, they are clustered around stars in parsec-sized haloes \citep{walker2017}. We therefore propose that the WIM seen in emission consists of D-WIM in the form of large numbers of tiny plasma clouds, clustered around stars. We note that a clustering scale of order a parsec, at a distance of $8\,{\rm kpc}$, corresponds to approximately 3~pixels of the \herschel\ PACS instrument, and may thus account for the intensity structure evident in the N$^+$ images shown by G15 (their figure 11).

An attractive feature of a stellar association is that the vertical distribution of the gas in the disk of the Galaxy, and thus the emission observed from that gas, is the same as that of the stars themselves. It is not necessary to fret about how the gas is supported in the gravitational field of the disk at heights $\sim1\,{\rm kpc}$. Gas associated with high mass stars, which have a small scale-height in the Galactic disk, yields emission predominantly at low Galactic latitudes, whereas gas associated with low mass stars contributes to the emission over a much larger vertical range.

The picture we are proposing for the D-WIM has implications for the way in which it is energised. By analogy with the very similar plasma structures found in the Helix nebula, \citet{walker2017} argued that the extreme scattering plasma probably forms on the surfaces of tiny molecular clouds. If so, photoionisation and UV pumping of H$_2$ would presumably be the main processes responsible for energising the D-WIM.  That implies two consequences that are in accord with what is known about the WIM in emission. First, massive stars are highly luminous, so if the D-WIM is energised by UV photons then its emissions ought to be concentrated in a thin disk, with weaker emission extending out to heights $\sim1\,{\rm kpc}$, as is observed (e.g. figure 4). Secondly, the ratio of ionising photons to sub-ionising UV photons is a strong function of the stellar photospheric temperature. As photoionisation injects only a small amount of heat per ionisation, whereas the opposite is true for UV pumping of H$_2$, there is an immediate expectation that cooler stellar photospheres should produce hotter plasma. As noted above, low mass stars (which have cooler photospheres) have larger scale heights in the Galactic disk, implying a systematic increase in plasma temperature as one moves away from the plane --- in agreement with what is inferred from the optical emission line data \citep{haffner2009}.

\section{Conclusions}
The ionised gas discovered by \citet{goldsmith2015} is so dense and highly pressured that it makes sense to identify it as a distinct component in its own right --- the D-WIM. It forms a geometrically thin disk in the inner Galaxy, and that disk probably exhibits strong spiral structure. D-WIM makes a substantial contribution to the observed pulsar dispersion measures in the inner Galactic plane, albeit with large point-to-point variations, and may dominate the DMs on some of these lines-of-sight. At high Galactic latitude, by contrast, it makes a negligible contribution to pulse dispersion.

The D-WIM dominates the scattering measure and the emission measure towards the inner Galactic plane, and probably contributes substantially to the observed emissions all over the sky. It is also a natural candidate to explain the extreme radio-wave scattering phenomena that are sometimes seen in pulsars and quasars. Drawing on what is known about the scattering material then leads to a picture of the D-WIM in which large numbers of tiny ($100\,{\rm AU}$) clouds are clustered in parsec-sized regions around stars. The velocity dispersion of the stars themselves provides vertical support against the gravitational potential of the disk.

Several pieces of evidence suggest that the D-WIM is too hot for a pure UV photoionisation model, and within a collisional ionisation equilibrium model we prefer a temperature of $19{,}000\,{\rm K}$ (inclusion of charge transfer reactions in the model would likely lower our preferred temperature, see Appendix A). In the context of a circumstellar interpretation of the D-WIM, an important heat source for the plasma may be UV pumping of H$_2$. This would mean that hotter plasma would be associated with cooler photospheres, leading to a correlation between the D-WIM temperature and height above the Galactic plane.

\section*{Acknowledgements}
{\it CHIANTI\/} is a collaborative project involving George Mason University, the University of Michigan (USA) and the University of Cambridge (UK). We benefitted from Ken Dere's help with {\it ChiantiPy\/}, from Dimitra Rigopoulou's perspective on the FIR lines, and from Dick Manchester's advice on pulsar DMs and the manuscript more broadly --- thanks Dick, Dimitra and Ken. MAW thanks Oxford Astrophysics for hospitality. MG thanks the Commonwealth Scholar Commission for DPhil funding and Manly Astrophysics and SKA SA for financial support and hospitality.  

\bibliographystyle{mnras}

\appendix
\section{Charge exchange}
As noted in the Introduction, the \chianti\ atomic database employed in this paper does not include charge exchange processes. Even if we consider only reactions of the form $A+B^+ \leftrightarrow A^++B$ (i.e. reactions between neutrals and singly charged ions) there are hundreds of processes to account for amongst the 30 abundant elements characterised in \chianti, so incorporating charge exchange processes is a challenging task. Many of the requisite rate coefficients are unknown \citep[e.g.][]{stancil2001}. However, given the rate coefficient for a particular reaction it is straightforward to determine the corresponding reaction rates under the conditions encountered in our model plasma, and at the suggestion of the Referee we have done that for the reaction $N+H^+ \leftrightarrow N^++H$. Figure A1 shows the results of that calculation for two different evaluations of the rate coefficients, from \citet{kingdon1996} and \citet{lin2005}. It is clear that at low temperatures the  charge exchange rates far exceed the corresponding total ionisation/recombination rates for nitrogen for all processes included in \chianti, so we expect that the collisional ionisation equilibrium solutions would be significantly modified if charge exchange were accounted for.

Because of the large number of charge exchange reactions, it is difficult to anticipate how our collisional ionisation equilibria would appear if charge exchange were incorporated in the model. However, given that our current model predicts the rate of $N+H^+ \rightarrow N^++H$ to exceed the rate of $N+H^+ \leftarrow N^++H$, at low temperatures, it seems likely that the ratio $N^+/N$ would peak at a lower temperature than our current model. In turn that is likely to lead to a decrease in the preferred temperatures when matching to data.

\begin{figure}
\includegraphics[width=85mm]{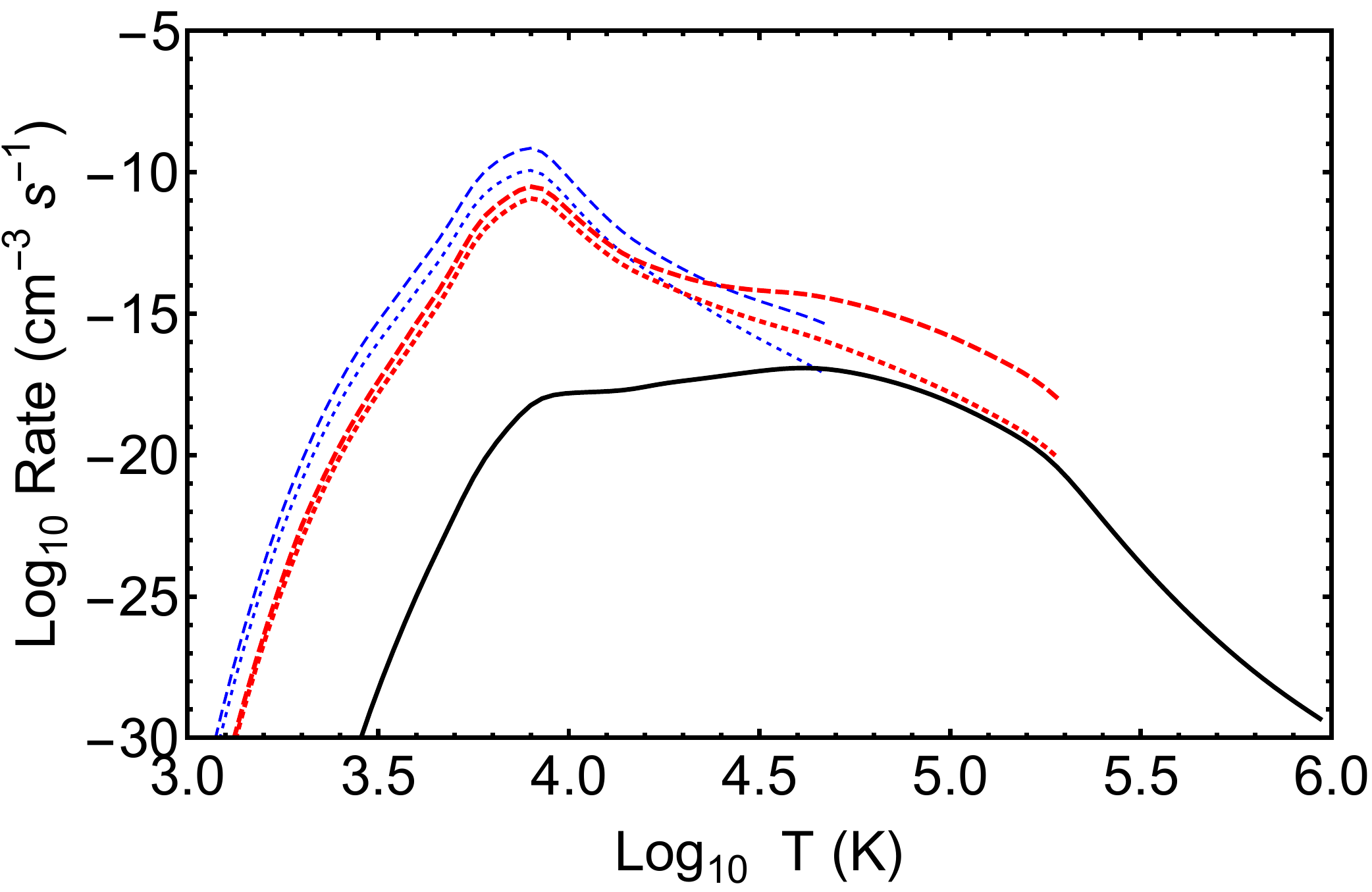}
\vskip-0.1truecm
\caption{Reaction rates in our model plasma, as a function of temperature, for (i) recombination of $N^+$ for all processes included in the \chianti\ database (black line), (ii) $N+H^+ \rightarrow N^++H$ (dashed lines), and (iii) $N+H^+ \leftarrow N^++H$ (dotted lines). For the charge exchange reactions (ii) and (iii) we show rates from \citet{kingdon1996} (thin, blue) and \citet{lin2005} (red). At low temperatures the charge exchange reactions clearly dominate.}
\label{fig:chargeexchange}
\end{figure}

\label{lastpage}
\end{document}